\documentclass[letterpaper,12pt,final,3p,times,longbibliography]{elsarticle}
\usepackage{graphicx,epsfig,url,lineno}
\usepackage{amssymb,fontenc,times,mathptmx}
\usepackage{wrapfig,rotating}
\usepackage[usenames]{color}
\usepackage{subcaption}
\usepackage{siunitx}
\usepackage[colorlinks = true,
            linkcolor = black,
            urlcolor  = blue,
            citecolor = blue,
            anchorcolor = blue]{hyperref}
\usepackage{tikz}
\usepackage{amsfonts}
\usepackage{systeme}

\usepackage{enumitem}
\usepackage{datatool-base}

\journal{Snowmass 2021 Proceedings}

\def \mtop {m_{\mathrm{t}}}
\def \mH {m_{\mathrm{H}}}

\def \mW {m_{\mathrm{W}}}
\def \mZ {m_{\mathrm{Z}}}
\def \GZ {\Gamma_{\mathrm{Z}}}

\def \WW {{\mathrm{W}^{+}} {\mathrm{W}^{-}} }

\def \ee {\mathrm{e}^{+}\mathrm{e}^{-}}
\def \electron {\mathrm{e}^{-}}
\def \positron {\mathrm{e}^{+}}
\def \Jpsi {J/\psi}
\def \KShort {\mathrm{K}^0_{\mathrm{S}}}

\def \mumu {\mu^+ \mu^-}
\def \mumug {\mu^+ \mu^- (\gamma)}
\def \eemmg { \mathrm{e}^{+}\mathrm{e}^{-} \rightarrow \mu^+ \mu^- (\gamma)}

\def \invfb {\mathrm{fb}^{-1}}
\def \invab {\mathrm{ab}^{-1}}

\def \sqrtsp {\sqrt{s}_{p}}
\def \sqrtsq {\sqrt{s}_{\psi}}
\def \sqrtsm {\sqrt{s}_{m}}

\def \Pel {P_{\mathrm{e^-}}}
\def \Ppos {P_{\mathrm{e^+}}}

\def \g1z {g_{1}^{\mathrm{Z}} }

\def \kspipi {\mathrm{K}^{0}_{\mathrm{S}} \to \pi^{+} \pi^{-}}
\def \Jmumu {\Jpsi \to \mumu}
\def \Lppi {\Lambda^{0} \to \mathrm{p} \pi^{-}}

\def \Ebm {E_{\mathrm{b}}^{-}}
\def \Ebp {E_{\mathrm{b}}^{+}}
\def \Eave {E_{\mathrm{ave}} }
\def \Ediff {\overline{\Delta E_{\mathrm{b}}}}

\def \pinibf { \mathrm{\mathbf{p}}_{\mathrm{ini}} }
\def \ponetwobf { \mathrm{\mathbf{p}}_{12} }
\def \ponebf { \mathrm{\mathbf{p}}_{1} }
\def \ptwobf { \mathrm{\mathbf{p}}_{2} }
\def \pthreebf { \mathrm{\mathbf{p}}_{3} }
\def \abf { \mathrm{\mathbf{a}}}
\def \bbf { \mathrm{\mathbf{b}}}
\def \Eini {E_{\mathrm{ini}}}
\def \ca {c_{\alpha}}
\def \sa {s_{\alpha}}
\def \ta {t_{\alpha}}
\def \p12mag {|\vec{p}_{12}|}

\def \Ep {E^{\prime}}

\begin{document}

\begin{frontmatter}

\title{{\Large{\bf Center-of-mass energy determination using \\ $\ee$ $\rightarrow$ $\mumug$ events 
at future $\ee$ colliders}}}

\author{Brendon Madison and Graham W. Wilson}
\address{Department of Physics and Astronomy, University of Kansas, \\
  Lawrence, KS 66045, USA}

\begin{abstract}
Methods for measuring the absolute 
center-of-mass energy, $\sqrt{s}$, and its 
distribution, are investigated for future $\ee$ Higgs-factory 
colliders using {\it in situ} $\ee$ collisions. 
We emphasize the potential of an estimator 
based on the measurement of muon momenta that we denote $\sqrtsp$.
It can be determined with high precision in $\ee \to \mumu (\gamma)$ events
while being sensitive to effects from beam energy 
spread, beamstrahlung, initial-state radiation (ISR), final-state 
radiation (FSR), crossing angle, and detector resolution.
The measurement precision is enabled by 
a high-precision low-mass tracker; the reported performance 
estimates are based on full simulation of the 
tracker response of the ILD detector concept for the ILC 
operating at $\sqrt{s}=250$~GeV.  
The underlying statistical 
precision is 1.9 ppm for 
a 2.0 $\invab$ dataset at ILC.
The ultimate utility will depend largely on how well one 
can calibrate and maintain the tracker momentum scale.
\end{abstract}

\end{frontmatter}

\begin{center}
{\it Submitted to the Proceedings of the US Community Study \\ on the Future of Particle Physics (Snowmass 2021)}
\end{center}

\section{Introduction}

Various $\ee$ collider concepts are under investigation 
as a potential future ``Higgs factory'' to explore in detail the 
physics of the Higgs boson, 
make precision measurements in the electroweak 
and top sectors, and to explore potential new physics beyond 
the Standard Model. This precision $\ee$
program is highly complementary to the 
continued exploitation in hadron collisions of the LHC that will be enabled with the HL-LHC. 
The $\ee$ collider concepts vary in 
maturity, cost, and readiness for construction and 
include linear collider (ILC, CLIC, $\mathrm{C}^{3}$, HELEN, and ReLiC) 
and circular collider (CEPC, FCC-ee, and CERC) concepts. 
More details and links to documentation are given 
in~\cite{Roser} and~\cite{Bagger}.
A key element that enables the physics program of a 
future $\ee$ collider is the control of the center-of-mass 
energy ($\sqrt{s}$) scale of the $\ee$ collisions.

Higgs boson physics can start 
to be thoroughly explored 
at $\ee$ colliders operating above 
the nominal ZH threshold 
of $\sqrt{s}=216.4$~GeV. 
In this energy regime, the classic technique 
of determining the circulating beam 
energies by measuring 
the electron/positron spin precession 
frequency by resonant depolarization 
is not expected to work for 
circular colliders due to 
the large beam energy spreads\footnote{In a circular collider, the synchrotron radiation component of 
the beam energy spread, $\sigma_{E}$, scales as $E_{\mathrm{b}}^2/\!\!\sqrt{\rho}$~\cite{Helm:1973xn},  
where $E_{\mathrm{b}}$ is the beam energy, and $\rho$ is the ring radius.
Based on experience from LEP, 
sufficient transverse polarization for 
the resonant depolarization technique can 
be obtained if the total beam 
energy spread is below about 55~MeV~\cite{Blondel:2019jmp}.},
and does 
not apply 
to linear colliders at 
any center-of-mass energy. 

Some of the legacies of lower-energy $\ee$ colliders, and prime examples 
of the utility of controlling the center-of-mass energy, 
are the precision determinations of the mass of 
the Z boson at the CERN LEP collider~\cite{LEP1} 
and the mass of the $\Jpsi$ particle at the Novosibirsk VEPP-4M collider~\cite{Anashin:2015rca}.
Center-of-mass energy scans resulted in  
measurements of the line-shape of 
the cross-section leading to knowledge of 
the Z mass to 23~ppm and the $\Jpsi$ mass to 1.9 ppm using 
resonant depolarization based beam energy calibration. 
However, already at LEP2, the resonant depolarization
technique was not feasible at the center-of-mass energies 
associated with $\WW$ production, and alternative 
techniques were needed. At the next generation 
high-energy $\ee$ collider we expect further 
marked improvements on measurements 
of the masses of the Z, W, and Higgs bosons and the top quark; 
in each case understanding the center-of-mass energy scale can 
be key to the ultimate precision. Prospects for precision energy 
calibration at the Z and WW threshold using resonant depolarization 
are described in an FCC-ee study~\cite{Blondel:2019jmp} in the context of a large circular collider.
Examples of threshold scan based measurement prospects at linear colliders
are described for the W mass in~\cite{Wilson:2016hne} 
and the top quark mass in~\cite{Seidel:2013sqa}. 
ILC can be operated at the Z-pole~\cite{Yokoya:2019rhx} and how well one can 
control the center-of-mass energy using techniques such as discussed here 
is a key question for such operation for physics at linear colliders.

In this paper we discuss two primary center-of-mass energy 
measurement techniques to measure the center-of-mass 
energy using collision events with an emphasis on dimuon events.
The first is an angles based estimator 
in radiative-return Z events, that we denote here as $\sqrtsq$.
This was first 
proposed in 1996~\cite{GWWMPI}, written up 
briefly in~\cite{Brinkmann:1997ws}, 
and used and applied to all fermion-pair 
channels at LEP2~\cite{ALEPH:2006cdc, OPAL:2004xxz, L3:2003vcr, DELPHI:2006ymr},
mostly as a cross-check of accelerator measurements.
A later study focused on linear colliders 
is documented in ~\cite{Hinze:2005kh}, \cite{Hinze:2005xt}.
The second is a momentum based estimator, denoted $\sqrtsp$,
inspired by earlier work by Barklow~\cite{Barklow} 
that was discussed in some detail in~\cite{GWW-LC2013}
and published in conference proceedings~\cite{Wilson:2020arh}.
The over-arching idea for both techniques is 
to use the kinematics of $\eemmg$ events and 
measurements of the final-state particles to measure
the distribution of the center-of-mass energy of collisions. 
The $\sqrtsq$ method relies on knowledge of $\mZ$ 
for the absolute scale, and the $\sqrtsp$ method requires 
excellent knowledge of the momentum scale of the tracker.
In addition, we have just noticed a third related method 
which we denote, $\sqrtsm$, based on mass and angles, that 
merits more investigation.
These deceptively simple techniques 
appear feasible for all future $\ee$ colliders over 
a large range of $\sqrt{s}$.  

Our recent work has focused on the 
$\sqrtsp$ estimator and is the main focus of this report, 
for reasons that will become clear. 
Studies are presented in 
the context of the ILC project~\cite{ILCInternationalDevelopmentTeam:2022izu} 
with the ILC accelerator operating initially at $\sqrt{s}=250$~GeV 
and using the ILD detector concept~\cite{ILDConceptGroup:2020sfq} 
to model the detector response.

We first describe in Section~\ref{sec:env} the accelerator environment 
and the consequent beam energy and center-of-mass energy distributions expected.
Here the focus is on characterizing the full energy peak region of 
the center-of-mass energy distribution that is sensitive to the absolute $\sqrt{s}$ scale.
In Section~\ref{sec:angles} we discuss the use of dimuons under simplifying 
3-body kinematics assumptions to reconstruct $\sqrt{s}$ by the three methods. 
Section~\ref{sec:sqrtsp} adds more realism to the application of the $\sqrtsp$ 
approach including explicit allowance for crossing angle, beam energy difference, 
and massive recoil system. In Section~\ref{sec:analysis} we assess the potential measurement 
precision of the new $\sqrtsp$ based methodology using full simulations of 
the ILD detector concept.
Finally we give an outlook and conclude.

\section{$\ee$ collider environment with focus on ILC}
\label{sec:env}
Currently established linear collider designs avoid arc-based 
synchrotron radiation losses that limit circular collider energy reach 
by being linear, but are single-pass{\footnote{Linear colliders based on energy recovery 
would operate in a different regime.}}, 
and in order to achieve very high luminosities per bunch crossing, 
need high bunch charges and nanometer sized beams. The intense electromagnetic fields of the 
colliding oppositely charged bunches lead to an increase in luminosity over the geometric luminosity 
due to the pinch effect, but also frequent energy loss especially 
from emission of energetic beamstrahlung photons. 
Beam-beam phenomena including beamstrahlung have been the subject 
of much study since linear colliders were first developed, as these impact 
accelerator, detector, and physics performance. 
A thorough introduction is given in~\cite{Yokoya:1991qz}.

A particular focus of past studies has been the development 
of methods to make measurements of the shape of 
the probability distribution of the center-of-mass 
energy using collision events. 
This is traditionally referred to as 
measurement of the luminosity spectrum, and 
involves measuring essentially what is akin to 
the structure function of each beam resulting 
from beam energy spread and beamstrahlung.
The studied technique proposed by Frary and Miller~\cite{Frary:1991aa} 
has been to use Bhabha scattering events which benefit 
from a substantial t-channel enhancement where the electron and positron 
are scattered at polar angles in the 
tracker acceptance (roughly $7^{\circ}$ for ILD). This allows 
precise angular measurements and reconstruction of the 
acollinearity angle or related variables and calorimetric measurements 
of the energies.
The most recent extensive study along these lines is 
described in~\cite{Poss:2013oea} as applied 
to CLIC at 3~TeV. This detailed paper has many references 
to the earlier literature including studies 
done previously for ILC at $\sqrt{s}=500$~GeV by Sailer (one of the authors). 
The expected calorimetric energy resolution 
is typically 2\% for 125~GeV beam energy electrons. 
This contrasts with 
approximately 0.14\% momentum 
resolution for\footnote{This momentum corresponds to that of each 
muon from Z decay in $\mathrm{Z} \gamma$ events 
at $\sqrt{s}=250$~GeV for the case of center-of-momentum decay 
angle of $\cos{\theta^{*}} = 0$.} 70.8~GeV muons with 
a $p_{T}$ of 45.6~GeV.

Details of the ILC accelerator are described 
elsewhere~\cite{Bambade:2019fyw} and references therein, 
but of particular relevance here are 
the expected relative beam energy 
spread of 0.152\% for positrons and 0.190\% for 
electrons\footnote{In the baseline design, 
the energy spread for electrons is larger than 
that for positrons because the electron beam 
is used for undulator-based positron production.} 
for operation at $\sqrt{s}=250$~GeV. Other parameters 
of interest are the horizontal crossing angle, $\alpha$, of 
\SI{14}{\milli\radian}, the bunch length of 
\SI{300}{\micro\metre},
and the luminosity-weighted 
average beamstrahlung energy loss of each beam particle\footnote{Standard 
accelerator parameters quote the beamstrahlung energy loss 
per beam particle independent of whether the particle contributes significantly 
to the luminosity. This is more relevant 
for spent beam handling and downstream diagnostics than physics  
and amounts to 2.6\% for ILC250.}of 1.1\%. These lead to 
a quoted 73\% of the luminosity being at center-of-mass energies 
exceeding 99\% of the nominal 250~GeV center-of-mass energy.

\begin{figure}[!htbp]
\centering
\includegraphics[height=0.4\textheight]{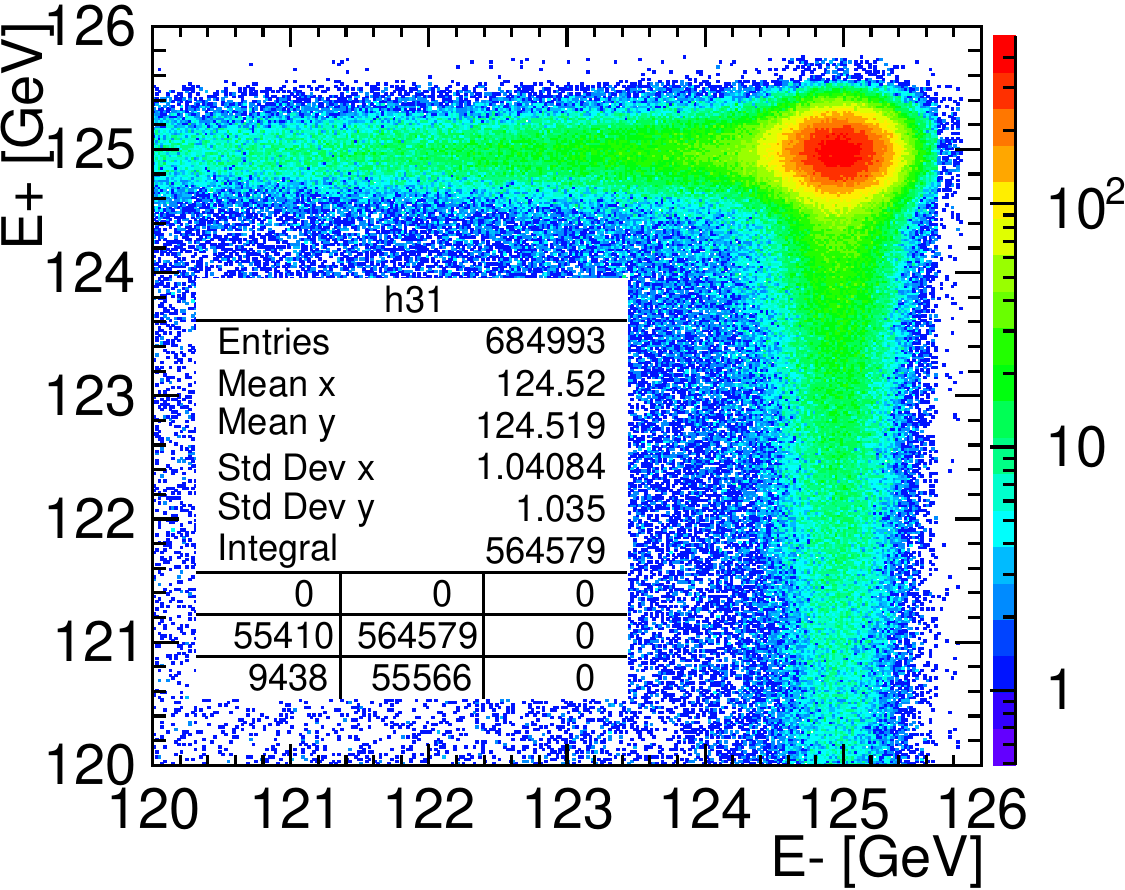}
\caption[]{\small \sl 
Distribution of the positron and electron beam energies at generator level 
for $\eemmg$ events. 
The simulation includes the effects of beam energy spread and beamstrahlung.
}
\label{fig:2dplot}
\end{figure}

We show in Figure~\ref{fig:2dplot}, the distribution at generator level 
of the positron beam energy ($\Ebp$) vs the electron beam energy ($\Ebm$), 
for events from the $\eemmg$ process as modeled using the WHIZARD 
physics event generator~\cite{Kilian:2007gr},\cite{Moretti:2001zz} for ILC 
operating at $\sqrt{s}=250$~GeV\footnote{The standard ILC run scenario at $\sqrt{s}=250$~GeV 
totals 2.0~$\invab$ for four different beam polarization configurations with 80/30\% electron/positron polarization. The figures used as illustration use $\eemmg$ with an integrated luminosity of 0.1~$\invfb$ 
for $P(\electron)=-0.8$ and $P(\positron)=+0.3$.} .
This includes effects of beam energy spread and beamstrahlung as modeled 
using the Guinea-Pig beam-beam simulation~\cite{Schulte:1997nga} and 
represented numerically using the 
CIRCE2 interface of WHIZARD that builds on the work of~\cite{Ohl:1996fi}.
This distribution illustrates the modeled uncorrelated bivariate 
Gaussian ``peak'' region, two ``arm'' regions where one of the 
beams undergoes significant energy loss from beamstrahlung, 
and a less populated ``body'' region where both beams have 
significant beamstrahlung energy loss. Here, 82\% of events result 
from interactions where the energy of both beam particles exceeds 120~GeV. 

\begin{figure}[!htbp]
\centering
\includegraphics[height=0.4\textheight]{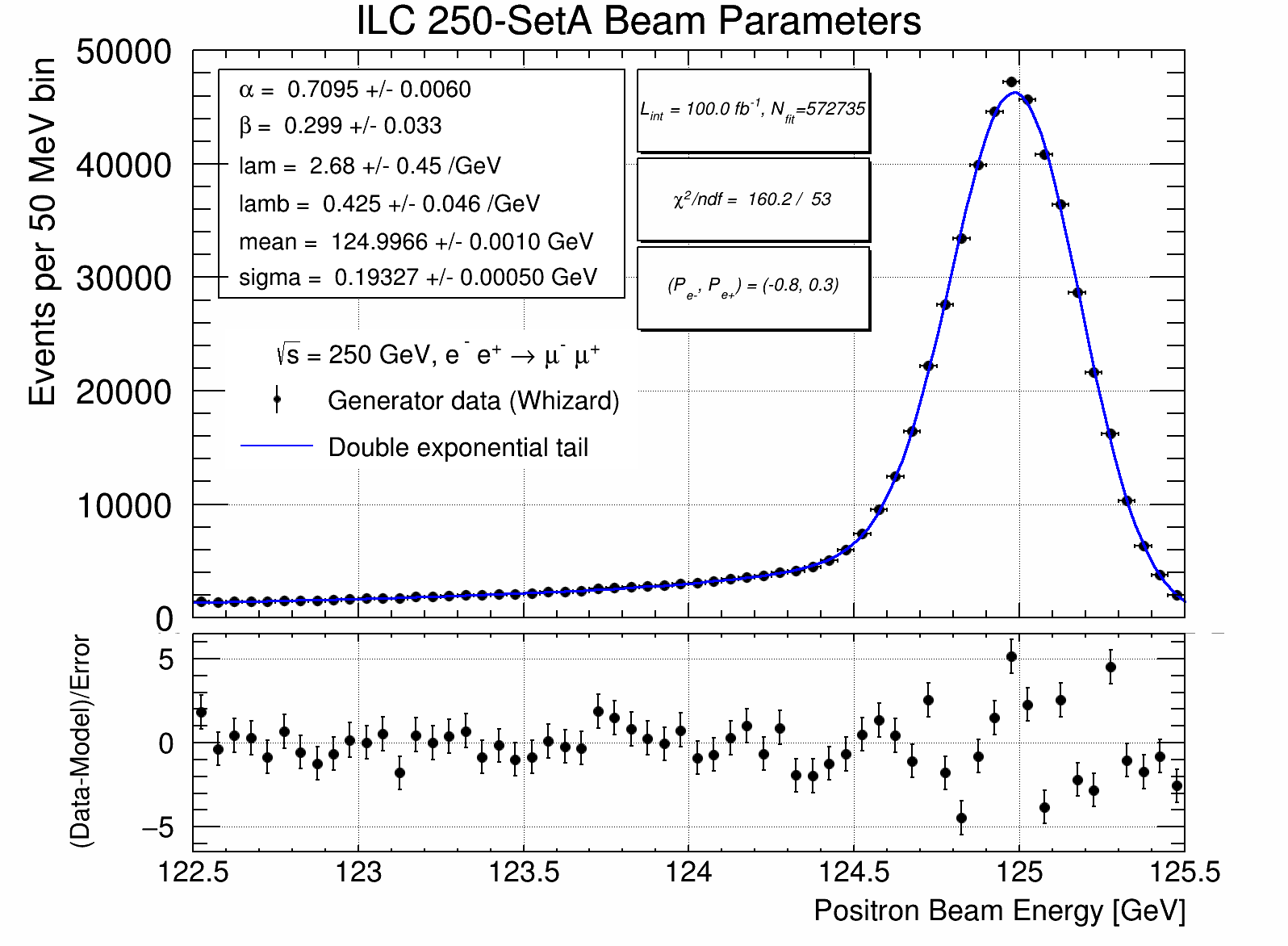}
\caption[]{\small \sl 
Distribution of the positron beam energy at generator level 
for $\eemmg$ events with super-imposed 6-parameter fit. 
The simulation includes the effects of beam energy spread and beamstrahlung.
Fit model details are discussed in~\ref{app:fits}.
}
\label{fig:positron}
\end{figure}

\begin{figure}[!htbp]
\centering
\includegraphics[height=0.4\textheight]{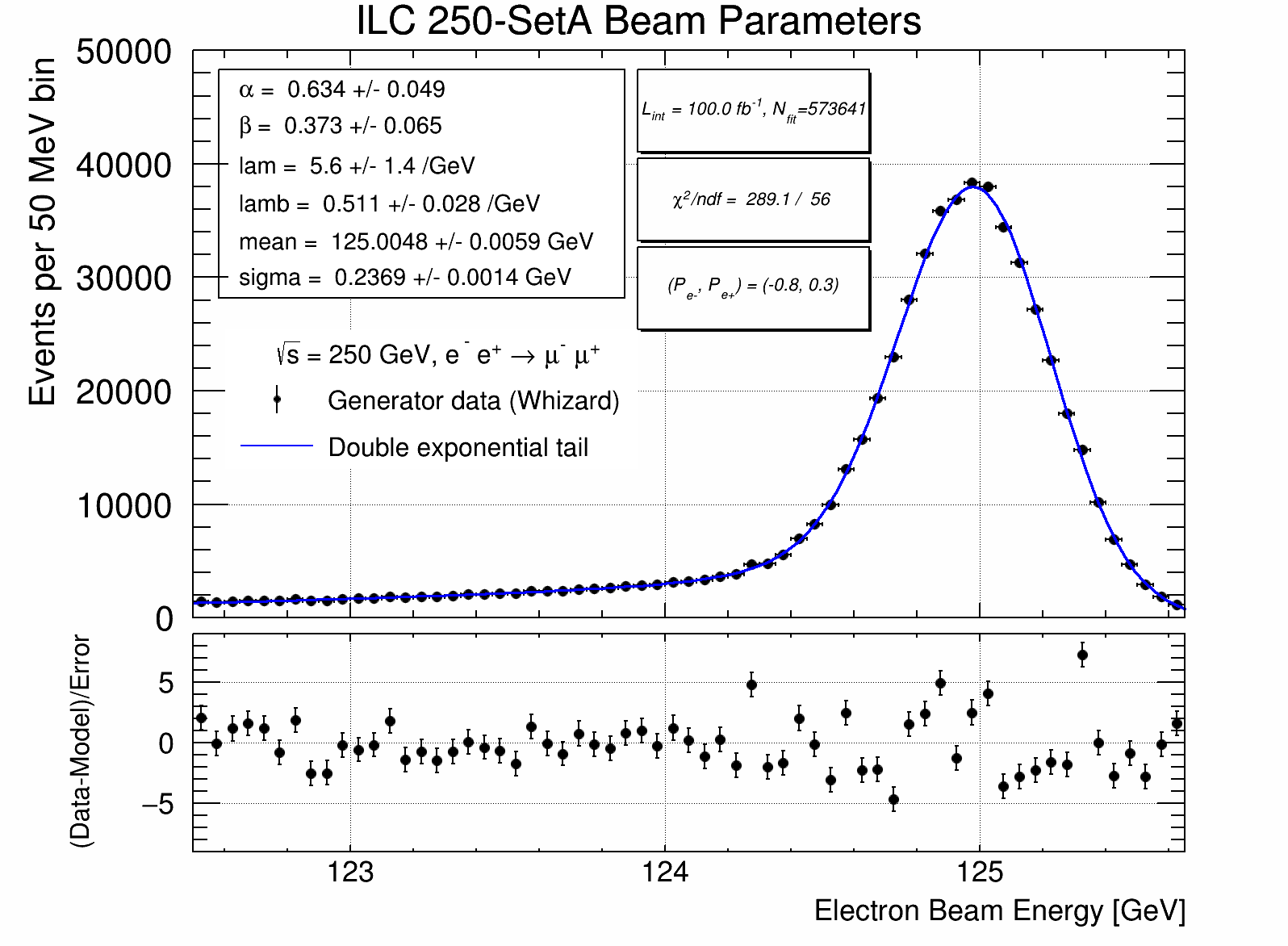}
\caption[]{\small \sl 
Distribution of the electron beam energy at generator level 
for $\eemmg$ events with super-imposed 6-parameter fit. 
The simulation includes the effects of beam energy spread and beamstrahlung.
}
\label{fig:electron}
\end{figure}

\begin{figure}[!htbp]
\centering
\includegraphics[height=0.4\textheight]{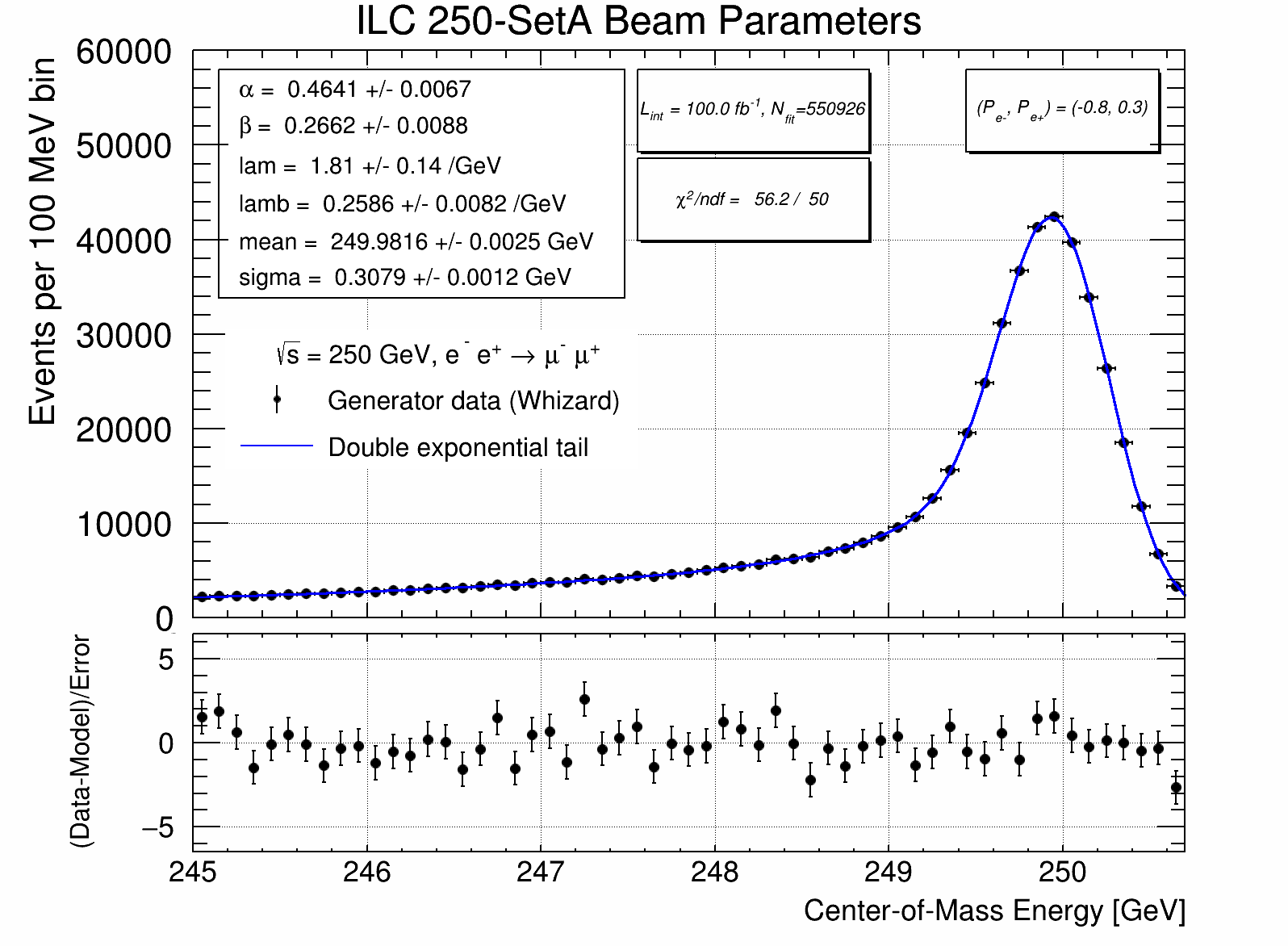}
\caption[]{\small \sl 
Generator level $\sqrt{s}$ distribution for $\eemmg$ events 
with super-imposed 6-parameter fit.
}
\label{fig:sqrtsee}
\end{figure}

\begin{figure}[!htbp]
\centering
\includegraphics[height=0.4\textheight]{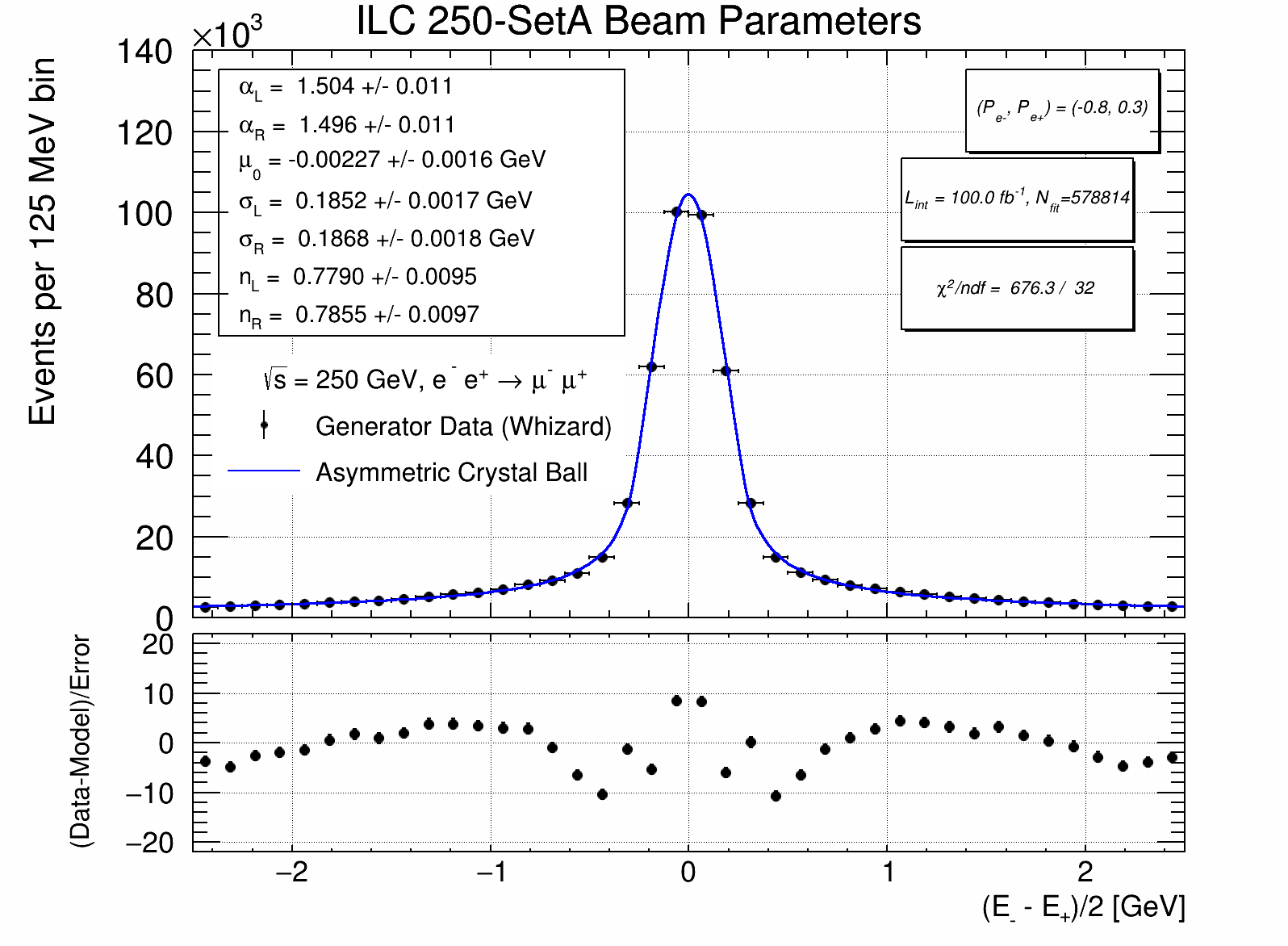}
\caption[]{\small \sl 
Generator level $\Ediff$ distribution 
for $\eemmg$ events. We also super-imposed a
two-sided Crystal Ball fit for illustration.
We have used the general fit parametrization 
that allows for non-equal parameters on the left and right of 
the location parameter, $\mu_{0}$. Clearly in this case the observations 
are consistent with being symmetric.
Note that for our current introductory purposes we are not concerned  
with parametrizing this distribution accurately. 

}
\label{fig:ediff}
\end{figure}

Also shown in Figures~\ref{fig:positron},\ref{fig:electron},\ref{fig:sqrtsee},\ref{fig:ediff} are the 1-d distributions at generator level 
of $\Ebp$, $\Ebm$, the center-of-mass energy, $\sqrt{s}$, and the distribution 
of ($\Ebm - \Ebp$)/2 for the same $\eemmg$ events. 
The distributions have superimposed empirical fit models. 
Fit models are detailed in~\ref{app:fits}.


We see that the fitted Gaussian width parameters are roughly 
consistent with the expectations of 0.152\% (positron) 
and 0.190\% (electron). These beam energy fits don't fit particularly well. 
Besides the possibility that the empirical model is inadequate, 
it is thought that there are at least two main underlying limitations in 
the degree of sophistication of the beam spectrum simulations that 
likely play a role in the poor fits. Firstly, the input beam 
energy distributions are truncated at about $\pm3\sigma$. 
Secondly, the input beam energy distributions were sampled from 
a limited set of 80,000 particles per beam, 
so some of these 
particles have been used more than once. 
We have mitigated the effect of the 
former for now by limiting the maximum 
energy considered in these and other fits.

By contrast the fit to the center-of-mass energy distribution (Figure~\ref{fig:sqrtsee}) 
is very acceptable. It illustrates a fitted center-of-mass energy 
spread of 0.308~GeV namely, 0.123\%, consistent with the 
expectation of 0.122\%. Figure~\ref{fig:ediff} shows that the modeled beam energy 
difference distribution is reasonably consistent with being symmetric. 
The overall width of the fitted central peak is about 22\% larger than expected 
from simple Gaussian energy spread. This could 
be a consequence of the poor fit model, or maybe more likely a reflection 
that some of the 1-d peak in energy difference comes from 
the body of Figure~1 when both beams radiate beamstrahlung 
photons of similar energies. This distribution also makes clear 
that the electron and positron beam energies often differ by much more than an 
amount characteristic of simply the beam energy spread, leading 
to a mostly longitudinal boost to the actual collision 
in addition to the small horizontal transverse boost associated 
with the crossing angle. A key issue will be getting an experimental 
handle on these energy differences and it is expected that collision-based measurements of 
$\eemmg$ events will also assist in the luminosity spectrum measurement.

\section{Reconstructing $\sqrt{s}$ with muons}
\label{sec:angles}



\begin{figure}
\centering
\begin{subfigure}{.5\textwidth}
  \centering
  \includegraphics[width=1.0\linewidth]{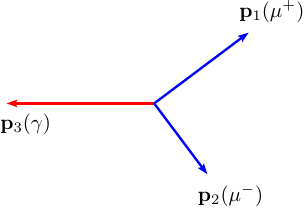}
  \label{fig:sub1}
\end{subfigure}%
\begin{subfigure}{.5\textwidth}
  \centering
  \includegraphics[width=1.0\linewidth]{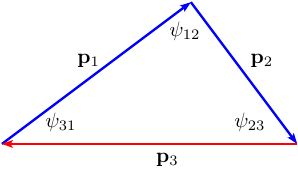}
  \label{fig:sub2}
\end{subfigure}
\caption[]{\small \sl 3-body $\mumu \gamma$ final state 
with corresponding triangle construction obtained from 3-momentum conservation and 
the defined  interior angles, $\psi_{12}$, $\psi_{23}$, $\psi_{31}$.}
\label{fig:triangle}
\end{figure}

Before getting into details, we will present the basic underlying principles of 
the two methods. They were both initially developed very simplistically 
with the underlying assumption that the lab. system is the center-of-momentum system. 
Three-body kinematics are in fact rather special. When one 
has a final state consisting of three particles as in
\begin{equation}
\ee \to {\mu_{1}}\!^{+}  {\mu_{2}}\!^{-} \gamma \, _{3}  \; \text{,}
\end{equation}
under the assumption of 3-momentum conservation, and being in the center-of-momentum system, one can write 
the 3-momentum conservation equation, namely,
\begin{equation}
\mathrm{\mathbf{p}_{1}}
+\mathrm{\mathbf{p}_{2}}
+\mathrm{\mathbf{p}_{3}}
= \mathrm{\mathbf{0}} \; \text{.}
\label{eqn:triangle}
\end{equation}
One way of looking at this is to view this 
as a triangle as illustrated in Figure~\ref{fig:triangle} with side lengths given by the 
magnitudes of each 3-momentum, $p_{1}$, $p_{2}$, $p_{3}$, 
and interior angles of the triangle, denoted, 
$\psi_{12}$, $\psi_{23}$, and $\psi_{31}$, where these necessarily satisfy, $\psi_{12} + \psi_{23} + \psi_{31} = \pi$.
Each interior angle is the supplementary angle to the opening angle between each particle pair. 
The three opening angles can be calculated simply from the 3-vector scalar product, so we have  
\begin{equation}
\psi_{12} \equiv \pi - \arccos\left( \frac{ \mathrm{\mathbf{p}_{1}} \cdot \mathrm{\mathbf{p}_{2}}    }{p_{1} \; p_{2}} \right) ,
\psi_{23} \equiv \pi - \arccos\left( \frac{ \mathrm{\mathbf{p}_{2}} \cdot \mathrm{\mathbf{p}_{3}}    }{p_{2} \; p_{3}} \right) ,
\psi_{31} \equiv \pi - \arccos\left( \frac{ \mathrm{\mathbf{p}_{3}} \cdot \mathrm{\mathbf{p}_{1}}    }{p_{3} \; p_{1}} \right) \; \text{.}
\end{equation}
Given that we now have a triangle, we can apply a relationship between 
the sides and angles of the triangle, namely the triangle sine rule, 
resulting in,
\begin{equation}
    \frac{p_{1}}{\sin{\psi_{23}}} 
  =  \frac{p_{2}}{\sin{\psi_{31}}} 
  =  \frac{p_{3}}{\sin{\psi_{12}}} \; \text{.}
\end{equation}
We also define the center-of-mass 
scaled photon energy, $x_{\gamma}$. For this 3-body case 
this will satisfy
\begin{equation}
x_{\gamma} \equiv \frac{2 E_{3}^{(*)}}{\sqrt{s}} = 1 - \frac{M_{12}^2}{s} \; \text{,}
\end{equation}
where $M_{12}$ is the dimuon mass, $E_{3}$ is the photon energy,  
and we note with the $^{(*)}$ notation that the photon energy 
needs to be evaluated in the center-of-momentum frame 
in general (as is the case here).
Now using the energy conservation equation
\begin{equation}
E_{1} + E_{2} + E_{3} = \sqrt{s} \; \text{,}
\label{eqn:consE}
\end{equation}
or rather the version of this where we neglect the muon mass (the photon mass is of course zero),
\begin{equation}
p_{1} + p_{2} + p_{3} = \sqrt{s} \; \text{,}  
\label{eqn:consE_approx}
\end{equation}
with the triangle sine rule expressions we can now write each momentum magnitude 
in terms of $\sqrt{s}$ and functions of the measured interior angles by eliminating the other momenta,
\begin{align}
\frac{p_{1}}{\sqrt{s}} = \frac{ \sin{\psi_{23}} }{ \sin{\psi_{12}} +  \sin{\psi_{23}} + \sin{\psi_{31}} } \; \text{,} \\ 
\frac{p_{2}}{\sqrt{s}} = \frac{ \sin{\psi_{31}} }{ \sin{\psi_{12}} +  \sin{\psi_{23}} + \sin{\psi_{31}} } \: \text{,} \\
\frac{p_{3}}{\sqrt{s}} = \frac{ \sin{\psi_{12}} }{ \sin{\psi_{12}} +  \sin{\psi_{23}} + \sin{\psi_{31}} } \; \text{.}
\end{align}

With these three expressions we have at hand 
a very useful tool. 
With knowledge of $\sqrt{s}$ one can apply each expression to the 
prediction of the muon momentum scale and the photon energy scale 
relying only on the precise angular measurements.

Now turning to the direct utility for this paper. The dimuon invariant mass squared is 
\begin{equation}
M_{12}^2 = 2 ( m^2 + E_1 E_2  - p_1 p_2 \cos{\theta_{12}} ) \; \text{,}
\label{eq:m1}
\end{equation}
where $\theta_{12}$ is the dimuon opening angle (ie. $\pi - \psi_{12} $). Again neglecting the muon mass, we get
\begin{equation}
M_{12}^2 \approx 2 p_1 p_2 ( 1 - \cos{\theta_{12}} ) = 2 p_1 p_2 ( 1 + \cos{\psi_{12}} ) \; \text{,}
\label{eqn:m12approx}
\end{equation}
and we can finally write
\begin{equation}
    \frac{ M_{12}^2} {s} = \frac{ 2 p_1 p_2 ( 1 + \cos{\psi_{12}}  )}{ s } 
   = \frac{ 2 \sin{\psi_{23}}  \sin{\psi_{31}}  ( 1 + \cos{\psi_{12}}  )     }
          {  ( \sin{\psi_{12}} +  \sin{\psi_{23}} + \sin{\psi_{31}} )^2                                              } \; \text{,}
\label{eqn:ratio}
\end{equation}

There is scope to retain the fermion 
masses that were 
neglected in Equations~\ref{eqn:consE_approx} and~\ref{eqn:m12approx}. Such added 
complication does not seem warranted given that the correctable bias on $\sqrt{s}$ 
associated with this approximation for muons is only +2.9~ppm for $\mathrm{Z} \gamma$ with $M_{12}=\mZ$ 
events at $\sqrt{s}=250$~GeV. Unsurprisingly, the bias for electrons is negligible (+70~ppb). 
It is more relevant for tau leptons, where the 
bias associated with omitting the 
mass is +0.08\% (assuming perfect measurement of the tau lepton direction\footnote{
Of course in the tau lepton case the direction of its visible decay products does not coincide with 
the tau lepton direction given the accompanying neutrino(s) and this will broaden 
and potentially bias the angular reconstruction.}),
and hadronic Z decays.

\subsection{Angles method, $\sqrtsq$}
The angles method uses angular estimates of the square of 
the ratio of the di-fermion mass to the center of mass energy,
such as Equation~\ref{eqn:ratio}, in radiative return to the Z events, $\ee \to \mathrm{Z} \gamma$, 
where the di-fermion mass is close to the Z mass.
Given that $\mZ$ is well known, one can rewrite  this equation to construct 
an estimator for the center-of-mass energy, namely
\begin{equation}
\sqrtsq = \mZ \: \frac{ \sin{\psi_{12}} +  \sin{\psi_{23}} + \sin{\psi_{31}}  } 
            {   \sqrt{ 2 \sin{\psi_{23}}  \: \sin{\psi_{31}} \: ( 1 + \cos{\psi_{12}}  )  }  }  \; \text{,}
\end{equation}
where the fermion masses have been neglected, and one 
reconstructs the distribution of $\sqrtsq$ under the assumption that 
the di-muon mass is actually $\mZ$. 
A second-order polynomial parametrization 
for the statistical precision of 
this method vs $\sqrt{s}$ was reported in~\cite{Hinze:2005xt} for unpolarized beams 
for center-of-mass energies ranging from 250 to 1000~GeV. At $\sqrt{s}=250$~GeV, the statistical precision was found to be 28~MeV for 100~$\invfb$ 
corresponding to a relative 
statistical precision of 25 ppm for 2.0~$\invab$ (unpolarized).


A similar expression for $x_{\gamma}$ can be formed using 
these angles. Other authors have 
used different conventions and notations for the angular measurements involved. 
We prefer this form for several reasons. Firstly, it retains a symmetry 
amongst the three particle pairs. Secondly, one of the interior angles is not 
explicitly eliminated using the $\psi_{12} + \psi_{23} + \psi_{31} = \pi$ constraint, and so one 
can use this as a test of the planarity assumption. Lastly, much 
of the literature has focused on using 3-body events in which the (ISR) 
photon is not in fact detected, and is assumed to be collinear with one of the beams. 
In this case with the beam $z$-axis defining the overall polar angle, 
it is then easy to confuse the polar angles of the muons with the angular 
separation of the muons from the photon.

The main perceived advantage of this method is that only precision 
angular measurements are required\footnote{While it is indeed reasonable to expect 
better resolution on angle measurements than momentum measurements, the ultimate 
systematic uncertainties are what count, and in this case 
the ability to align/calibrate is essential.}. 
There are four major limitations to this method. 
\begin{enumerate}
\item It requires $\mathrm{Z} \gamma$ events. So it does not make 
use of full energy $\ee \to \mumu$ events and will not be useful at $\sqrt{s} \approx \mZ$.
\item It relies on knowledge of $\mZ$. Currently this is 23~ppm. 
This will not be a limiting factor for anticipated measurements of $\mtop$ and $\mH$, 
but current knowledge is a limiting factor for much improved measurements of $\mW$ and especially $\mZ$.
On the other hand if $\mZ$ were to be markedly improved for example 
by dedicated running at the Z-pole as envisaged 
for both FCC-ee~\cite{Blondel:2019jmp} and ILC~\cite{Yokoya:2019rhx}, 
this would be less of a factor for the utility of this method.
\item The intrinsic precision per $\mathrm{Z} \gamma$ event is reduced by the sizable contribution to 
the event-by-event estimates from the intrinsic width of the Z. The effective resolution per event on the center-of-mass energy from this source is $\GZ/(\sqrt{2} \mZ)$ 
which equates to 1.9\%.
\item At high center-of-mass energies, the large boost associated with $\mathrm{Z} \gamma$ events 
means that it is more and more difficult for each decay muon to be retained within 
the detector acceptance. Even at just $\sqrt{s}=500$~GeV both muons will  
have polar angles within $21^{\circ}$ 
of the beam axis. 

\end{enumerate}

\subsection{Momentum method, $\sqrtsp$}
The overarching idea here is to just use the muon momenta, even if a photon is detected.
For the simplifying assumption that the lab. system is the center-of-momentum system, 
independent of whether the $\eemmg$ events are close to full energy $\ee \to \mumu$ like  
events with no obvious evidence for photon radiation, 
or $\ee \to \mumu \gamma$, one can construct an estimator of the center-of-mass energy 
under the assumption that the 
third particle system (the single photon in the previous discussion) is massless, by simply 
noting that Equation~\ref{eqn:triangle} can be re-expressed as 

\begin{equation}
\mathrm{\mathbf{p}_{3}} = -
 (\mathrm{\mathbf{p}_{1}}
+\mathrm{\mathbf{p}_{2}}) \; \text{,}
\end{equation}
and so the energy conservation equation (\ref{eqn:consE}), can be re-written simply as
\begin{equation}
    E_{1} + E_{2} + |\mathrm{\mathbf{p}_{1}}
+\mathrm{\mathbf{p}_{2}} | = \sqrt{s} \; \text{,}
\end{equation}
retaining the fermion masses.
Given that we are dealing with muons this results in the following estimator, that we denote, $\sqrtsp$, 
where 
the equation has been rewritten to make the momentum dependence completely explicit for
\begin{equation}
\sqrtsp = \sqrt{p_{1}^2 + m_{\mu}^2} + \sqrt{p_{2}^2 + m_{\mu}^2}  +  |\mathrm{\mathbf{p}_{1}}
+\mathrm{\mathbf{p}_{2}} | \; \text{.}
\label{eqn:sqrtsp}
\end{equation}
The main advantage of this technique over the angles method 
is much better intrinsic precision per event presuming an ILC-like tracker. 
It does rely on knowing the tracker momentum scale with high precision. It 
also benefits from higher numbers of available events as it does not 
require $\mathrm{Z} \gamma$ events. 
One area of concern is that for events 
that are essentially 2-body events, namely, $\ee \to \mumu$, the 
reconstruction method will always 
add in a third particle, a supposed photon, to balance the observed 
non-zero dimuon momentum that may arise simply from detector resolution.
As we will see the statistical 
precision of this method for muons is an order of magnitude better than the angles method.

\subsection{Mass method, $\sqrtsm$}
Given that we really do not want to necessarily assume the Z mass, but we have available the 
directly measured dimuon mass, with mass resolution much better than $\GZ$, we can 
also rewrite Equation~\ref{eqn:ratio} simply as 
\begin{equation}
\sqrtsm = M_{12} \: \frac{ \sin{\psi_{12}} +  \sin{\psi_{23}} + \sin{\psi_{31}}  } 
            {   \sqrt{ 2 \sin{\psi_{23}} \: \sin{\psi_{31}} \: ( 1 + \cos{\psi_{12}}  )  }  } \; \text{,}
\end{equation}
where we measure the dimuon mass directly, including the mass terms\footnote{Or they 
could be excluded if more appropriate in this context.}, 
and use the angular measurements to infer the scaling to 
the center-of-mass energy for the radiated photon. 
This is hot off the press and it remains to be seen how complementary 
this approach is to $\sqrtsp$. 
It looks less overtly a pure momentum-based measurement and appears to have 
exchanged a momentum-scale problem for a mass-scale problem, but we suspect this is naive given that the 
angles do also enter into Equation~\ref{eqn:sqrtsp}.

\begin{figure}[t]
\centering
\includegraphics[height=0.4\textheight]{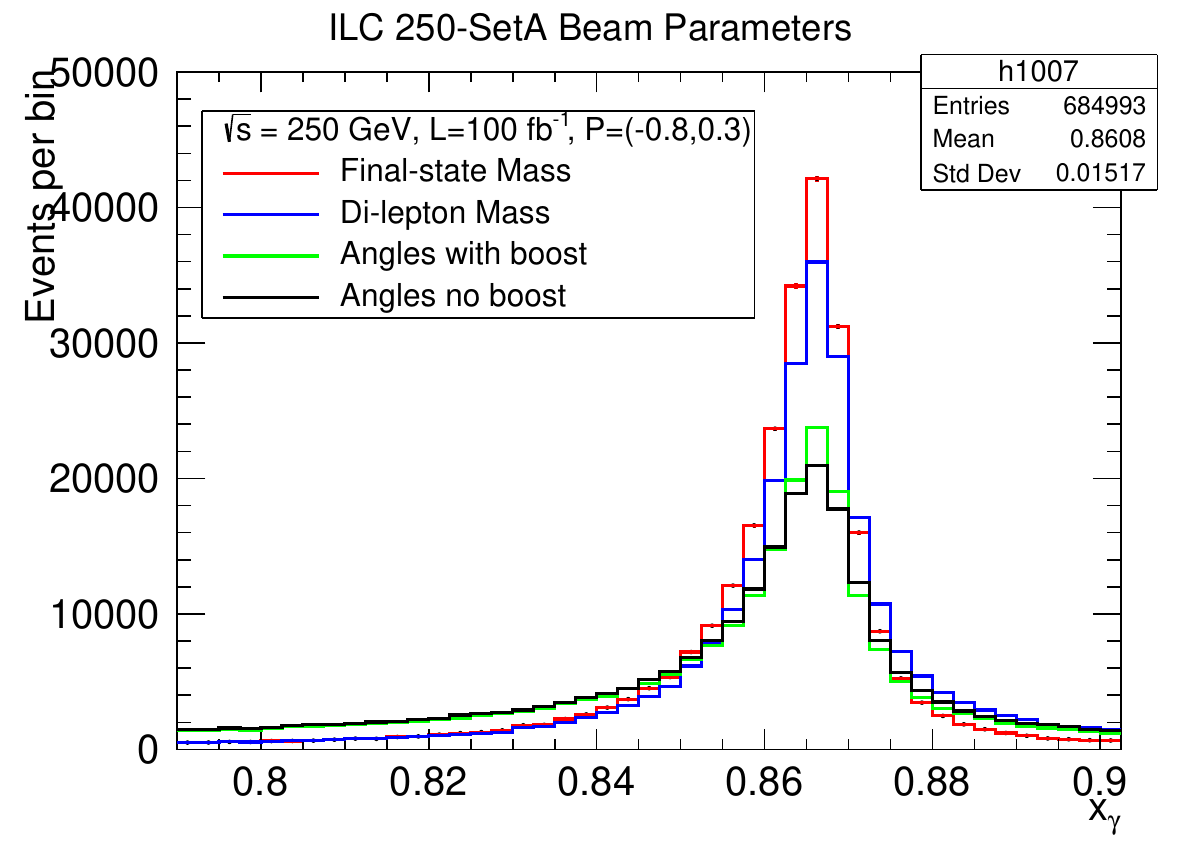}
\caption[]{\small \sl 
Comparison of different methods at generator 
level to reconstruct the peak in 
the center-of-mass scaled photon energy distribution of 
the assumed single 
photon in $\mathrm{Z} \gamma$ like dimuon events for 
$\sqrt{s}=250$~GeV. $x_{\gamma}$ is equivalent to $1 - M_{12}^{2}/s$. 
The black and green histograms use angular reconstruction. 
The blue histogram uses the dimuon mass, and for the red histogram the mass estimate 
additionally includes contributions from FSR photons.
}
\label{fig:sqrtsq}
\end{figure}

Some aspects of these methods are compared 
in Figure~\ref{fig:sqrtsq} based on generator level information in 
terms of the reconstructed $x_{\gamma}$ distributions 
either from muon angles 
or from knowledge of the mass of the dimuon and the actual center-of-mass energy. 
One expects a Breit-Wigner like peak centered 
at a photon energy of 108.4~GeV ($x_{\gamma} = 0.8670$) for $\sqrt{s}=250$~GeV.

For the angles based method it is assumed that the photon is collinear with the 
beam particle that best accommodates momentum balance and is undetected. The simulation 
includes crossing angle effects so in principle these should be corrected for 
before measuring/calculating angles. A backward horizontal boost\footnote{This technically makes the angles method momentum dependent - but a cursory examination indicates the effect is small.} 
from the lab of $\beta = \sin(\alpha/2)$ is used in the green curve so that the newly calculated 
angles are in the nominal center-of-momentum frame.
The black histogram neglects this correction. The blue and red 
histograms use the true 
dimuon mass directly, and the 
red histogram supplements this mass with any 
contributions from FSR photons as defined by the generator.
The blue and red histograms also use the true center-of-mass energy. 
The width of the red histogram simply reflects 
the underlying Z width. Clearly there is some room for 
improvement by including FSR photons in the Z decay system.
It is also apparent that the center-of-mass estimator with the angles-based method (green) will perform significantly worse than 
that implied by the intrinsic width of the Z alone unless more can 
be done to recuperate or correctly analyze the events that do not conform well 
to the assumptions.

\section{Realistic Kinematics for $\sqrtsp$ Estimate}
\label{sec:sqrtsp}
The main idea is again to use the kinematics of $\eemmg$ events and 
measurements of the final-state particles to measure
the distribution of the center-of-mass energy of collisions. The overall 
center-of-mass energy scale is provided by the tracker measurements of the muon momenta.
In the real world, we need to add more realism 
to the picture of what happens in the collisions. Each line of 
the following adds more realism, but is still in some aspects idealized:
\begin{enumerate}
    \item Nominal. Each beam is a delta function centered 
    at a particular beam energy, and the lab. frame is the center-of-momentum frame.
    \item Crossing Angle ($\alpha$). In practice beams will have a small net horizontal momentum 
    that leads to the detector (ie lab.) reference 
    frame never being the center-of-momentum frame.
    \item Beam energy spread (BES). Each beam naturally has an energy spread that reflects 
    the production, damping, transport, bunch compression, acceleration, and focusing 
    associated with preparation of the beam for delivery to the interaction point.
    A simplifying assumption is to use a Gaussian energy distribution. More realistic 
    models for ILC should include likely energy-$z$ correlations that can be modeled 
    with an end-to-end accelerator simulation, and may be constrained with beam diagnostics.
    Additionally each beamline is independent so there is no expectation that the two 
    beam energies are identical.
    \item BES + Beamstrahlung (BS). The collective interaction of the two beams leads to 
    radiation of photons referred to 
    as beamstrahlung photons from the beams, resulting in a beamstrahlung-reduced center-of-mass energy.
    \item BES + BS + Initial-state-radiation (ISR). All $\ee$ physics processes may have ISR, where the 
    invariant mass of the annihilating $\ee$ and the resulting particle system (excluding the ISR photon(s)) is reduced compared with case 3 due to the emitted ISR photon(s).
\end{enumerate}

We will be primarily concerned with evaluating the beamstrahlung-reduced center-of-mass energy. 
This is {\it after} beam energy spread and beamstrahlung radiation, but {\it before} emission of any 
ISR photons. We will allow for differences in the energy of each beam and 
for a beam crossing angle, $\alpha$, defined as the horizontal plane angle between the two beam lines.
For ILC $\alpha$ is 14~mrad.

\begin{figure}[!htbp]
\centering
\includegraphics[width=0.65\textwidth]{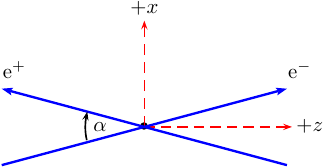}
\caption[]{\small \sl 
The $z$-axis is 
defined as the axis that bisects the outgoing electron beam-line axis
and the negative of the incoming positron beam-line axis. The $x$-axis is 
the horizontal axis in the direction of the net horizontal momentum of the initial state.
}
\label{fig:crossingangle}
\end{figure}

There are a number of formulations of the kinematic problem that differ in what are the 
assumptions, the measurements, and the setup. 
We focus on the system of 4-vectors comprising the $\ee$ 
after beamstrahlung emission as measured in the detector reference frame illustrated and defined in Figure~\ref{fig:crossingangle}.
Note that given the crossing angle, this reference frame is not the center-of-momentum frame.

Let's define the two beam energies (after beamstrahlung) 
as $\Ebm$ and $\Ebp$ for the electron and positron beam respectively, 
and define for simplicity, $\sa=\sin(\alpha/2)$, $\ca=\cos(\alpha/2)$ and $\ta=\tan(\alpha/2)$. This leads to an initial-state energy-momentum 4-vector $(E, \: p_{x}, p_{y}, p_{z})$ consisting of
\begin{flalign}
E &= \Ebm + \Ebp \; \text{,} \\
p_{x} &= (\Ebm + \Ebp) \: \sa \; \text{,}\\
p_{y} &= 0 \; \text{,} \\
p_{z} &= (\Ebm - \Ebp) \: \ca \; \text{,}
\end{flalign}
where we have neglected the electron mass.

The corresponding center-of-mass energy (the invariant mass of the initial-state 4-vector) is 
\begin{equation} \sqrt{s} = 2 \sqrt{\Ebm \Ebp} \: \ca \; \text{.}
\end{equation}

Hence if $\alpha$ is known, evaluation of the center-of-mass 
energy of this collision amounts to measuring the two beam energies. 
It is convenient to introduce the average of the two beam energies, $\Eave$,  
and half of the beam energy difference, $\Ediff$,
\begin{equation} \Eave \equiv \frac{\Ebm + \Ebp}{2} \; \text{,} \end{equation}
\begin{equation} \Ediff \equiv \frac{\Ebm - \Ebp}{2} \; \text{.} \end{equation}
With this notation 
\begin{equation} \sqrt{s} = 2 \sqrt{\Eave^{2} - (\Ediff)^{2}} \: \ca \; \text{,}
\label{eqn:sqrts}
\end{equation}
\begin{equation} p_{z} = 2 \; \Ediff \: \ca \; \text{,} 
\label{eqn:pz}
\end{equation}
and we can define the initial-state $\ee$ system 4-vector components as 
\begin{equation} (\Eini, \pinibf) \equiv 2 ( \Eave, \; \Eave \; \sa,\; 0, \;\Ediff \;\ca    )  \; \text{.}
\label{eqn:inistate}
\end{equation}
Now let's look at the final state of the $\eemmg$ process. 
We will denote the $\mu^{+}$ as particle 1, the $\mu^{-}$ as particle 2, and the rest-of-the event as 
system 3. In the simple case of $\ee \to \mumu$, with only two produced particles, 
the rest-of-the event is the null vector. 
In the case of one single photon, the rest-of-the-event is this one photon. 
In the case of multiple photons, the rest-of-the-event is this system of multiple photons, 
that may or may not have significant mass.
The photons are often ISR photons that tend to be at forward angles and can be undetectable.

We can write this final-state system 4-vector as
\begin{equation}  (E_{1} + E_{2} + E_{3}, \: \ponebf + \ptwobf + \pthreebf ) \; \text{.} \end{equation}
In general the rest-of-the-event system will not be fully detected and needs to be 
inferred using energy-momentum conservation or additional assumptions on especially its mass.
In full generality let's keep non-zero mass for the rest-of-the event mass, $M_{3}$.
Then proceeding to apply energy-momentum conservation by equating the above 
to the initial-state 4-vector we obtain 

\begin{equation} E_{1} + E_{2} + \sqrt{p_{3}^{2} + M_{3}^2} = 2 \; \Eave \; \text{,}  \label{eqn:E} \end{equation}
\begin{equation} \ponebf + \ptwobf + \pthreebf = \pinibf = 2( \; \Eave \: \sa, \;0 ,  \;\Ediff \: \ca ) \; \text{.}
\label{eqn:pinitial}
\end{equation} 

This represents the four equations of energy-momentum conservation and six unknowns, 
namely the three components of the rest-of-the-event 3-momentum, $\pthreebf$, 
the two beam energy related quantities, $\Eave$, $\Ediff$ and the 
rest-of-the-event mass $M_{3}$. 
Further progress needs additional assumptions\footnote{In prior expositions, we had needlessly assumed $M_{3}=0$ throughout.}.
A general way to approach the problem is to 
solve for $\Eave$ for various assumptions on $\Ediff$ and $M_{3}$. 
Specifically we will then focus on using the simplifying assumptions that $\Ediff = 0$ and $M_{3}=0$. 
The $\Ediff$ assumption is a poor one event-by-event for the $p_{z}$ conservation component as can be deduced from Figure~\ref{fig:ediff}.
In the following, we solve the general problem ($\Ediff$ and $M_{3}$ non-zero) and 
make no assumption on the direction of the rest-of-the event particles, apart from their total 
invariant mass.

We can write
\begin{equation} \pthreebf = \pinibf - (\ponebf + \ptwobf) \; \text{,} \end{equation} 
and using 
\begin{equation} | \abf - \bbf | = \sqrt{a^2 + b^2 - 2 \; \abf \cdot \bbf } \; \text{,} \end{equation} 
find
\begin{equation} |\pthreebf| =\sqrt{ p_{\mathrm{ini}}^2 + (p_{12})^2 - 4(p_{12}^{x}) \Eave \: \sa - 4 (p_{12}^z) \Ediff \: \ca } \; \text{.} 
\label{eqn:one}
\end{equation}
Now using equation~\ref{eqn:E} we can write,
\begin{equation} p_{3}^{2} = (2 \Eave - E_{1} - E_{2})^{2} - M_{3}^{2} \; \text{,} \end{equation}
and by equating this to the square of equation \ref{eqn:one}, we eliminate $\pthreebf$, 
\begin{equation} (2 \Eave - E_{1} - E_{2})^{2} - M_{3}^{2} = p_{\mathrm{ini}}^2 + (p_{12})^2 - 4(p_{12}^{x}) \Eave\: \sa - 4 (p_{12}^z) \Ediff \: \ca \; \text{.} \end{equation}  
Now inserting the components of $\pinibf$, we find, \\
\begin{equation} (2 \Eave - E_{12})^{2} - M_{3}^{2} = 4 \Eave^{2} \sa^2 + 4 \Ediff^{2} \ca^2 \\ 
                                       + (p_{12})^2 - 4(p_{12}^{x}) \Eave \sa - 4 (p_{12}^z) \Ediff \ca \; \text{.}
                                    \end{equation} \\
This is a quadratic
\begin{equation} 
A \Eave^2 + B \Eave + C = 0 \; \text{,}
\label{eqn:quadratic}
\end{equation}
in $\Eave$ with coefficients (after simplification) of 
\begin{flalign}
A &= \ca^2  \; \text{,} \\
B &= -E_{12} \; + \; p_{12}^{x} \: \sa \; \text{,} \\
C &= (M_{12}^{2} - M_{3}^{2})/4 \; + \; p_{12}^{z} \Ediff \: \ca \; - \; \Ediff^{2} \: \ca^2 \; \text{.} \end{flalign}

Based on this, there are three particular cases of interest which can be 
solved for $\Eave$ with in the first instance, $M_{3}=0$,
\begin{enumerate}
  \item Zero crossing angle, $\alpha=0$, and zero beam energy difference.
  \item Crossing angle and zero beam energy difference.
  \item Crossing angle and non-zero beam energy difference.
\end{enumerate}

The original studies led to the relationship 
\begin{equation} \sqrt{s} = E_{1} + E_{2} + |\ponetwobf| \; \text{,}  \end{equation} 
and this arises of course essentially trivially in the first case. 
Explicitly for this case, we have
\begin{equation} A = 1, B = -E_{12}, C= (M_{12}^2)/4 \; \text{,} \end{equation}
with a resulting discriminant of
\begin{equation} B^{2} - 4 A C = E_{12}^{2} - M_{12}^2 = p_{12}^2 \; \text{,} \end{equation}
and solutions of 
\begin{equation} \Eave = \frac{E_{12} \pm  |\ponetwobf|} {2} \; \text{,} \end{equation}
leading to 
\begin{equation} \sqrt{s} = 2 \; \Eave = E_{12} \pm |\ponetwobf| \; \text{.} \end{equation}
Only the positive solution is physical given that the $E_{3}$ contribution 
to the center-of-mass energy needs to be non-negative.

For cases, 2 and 3, the discriminant is always positive. It also 
seems that it is always the positive solution that 
yields the correct physical solution for $\Eave$. 
Given that the crossing angle is small, and the 
expected energy differences are relatively small, 
it is not unexpected that the more complete cases find 
solutions similar to the most approximate one (case 1). 
We have also re-derived the result of case 2 as a quadratic 
directly in terms of $\sqrt{s}$. This alternative formulation is described in~\ref{sec:alt} for completeness, and leads explicitly to the observation that the case 2 estimate, can be viewed simply as
\begin{equation}
\sqrtsp = E_{12}^{*}+   |\mathrm{\mathbf{p}}_{12}^{*}| \; \text{.} 
\end{equation}

\begin{figure}[!htbp]
\centering
\includegraphics[height=0.4\textheight]{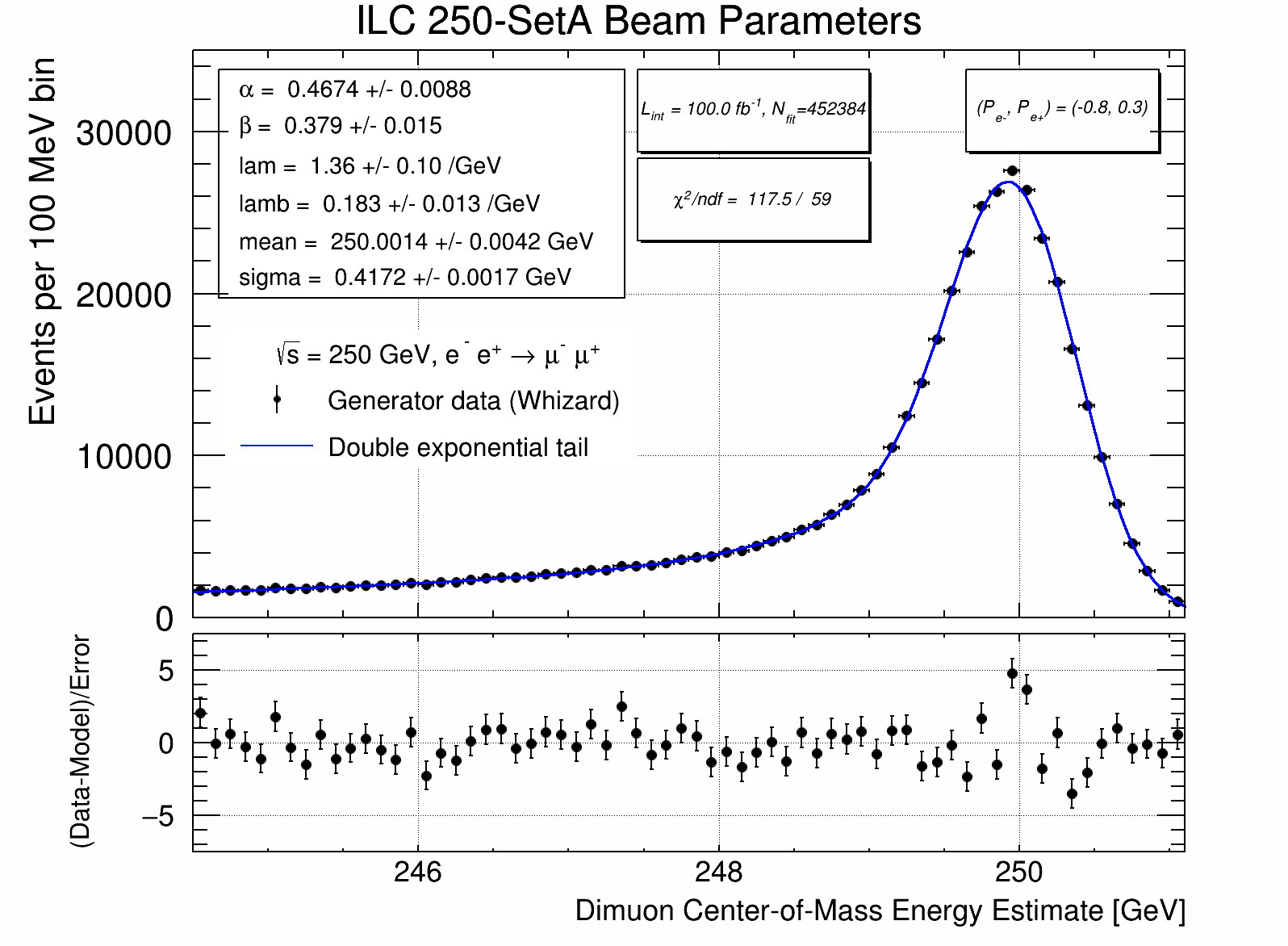}
\caption[]{\small \sl 
Generator level $\sqrtsp$ distribution evaluated with generator-level muons 
for $\eemmg$ events with super-imposed 6-parameter fit.
}
\label{fig:sqrtsp}
\end{figure}

\begin{figure}[!htbp]
\centering
\includegraphics[height=0.4\textheight]{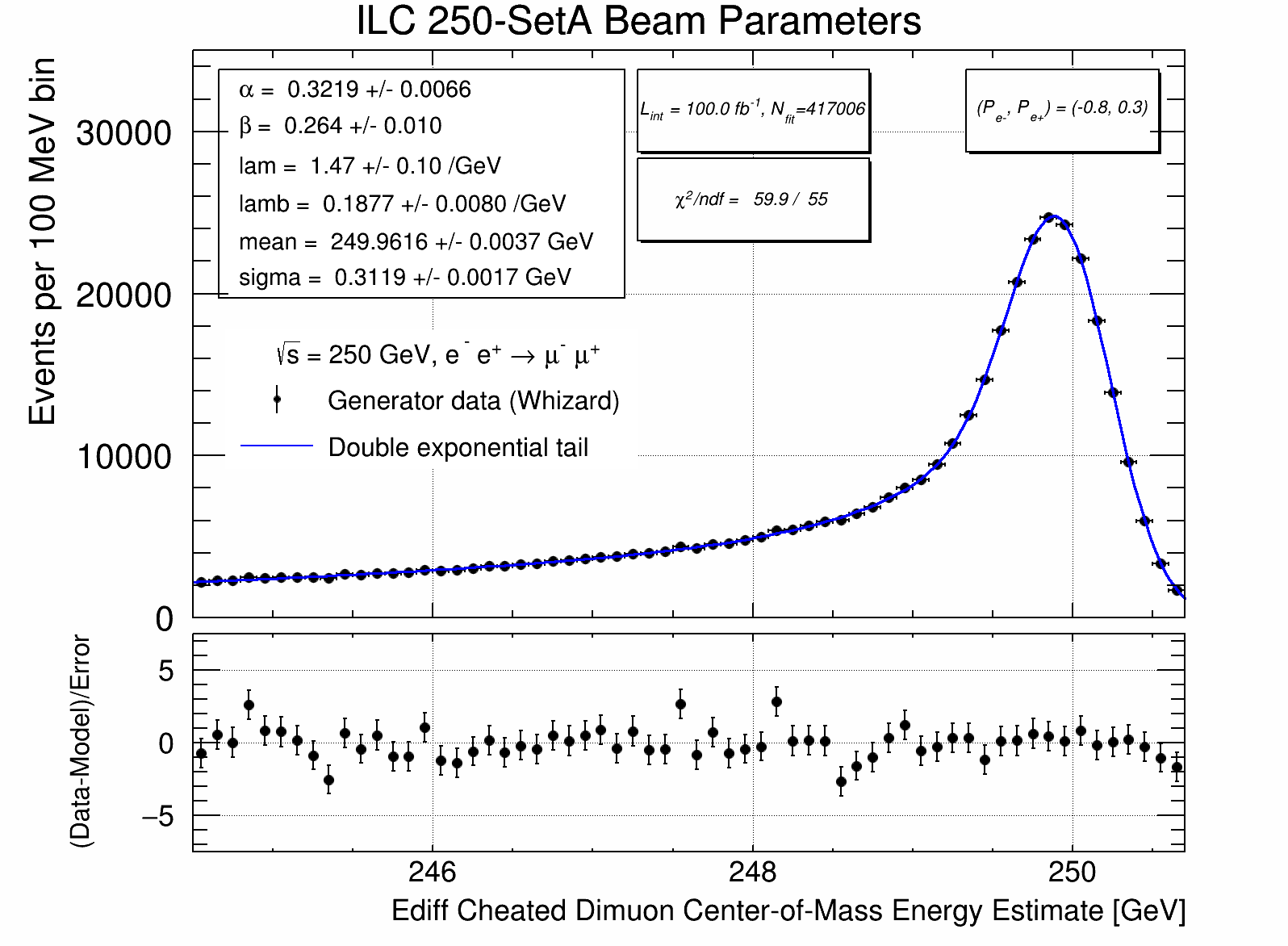}
\caption[]{\small \sl 
Generator level $\sqrtsp$ distribution evaluated with generator-level muons 
for $\eemmg$ events with super-imposed 6-parameter fit. In this case 
the true event-by-event value of the (halved) beam energy difference 
of Figure~\ref{fig:ediff} is used, rather than assuming that it is zero.
}
\label{fig:sqrtspcheated}
\end{figure}

\begin{figure}[!htbp]
\centering
\includegraphics[height=0.4\textheight]{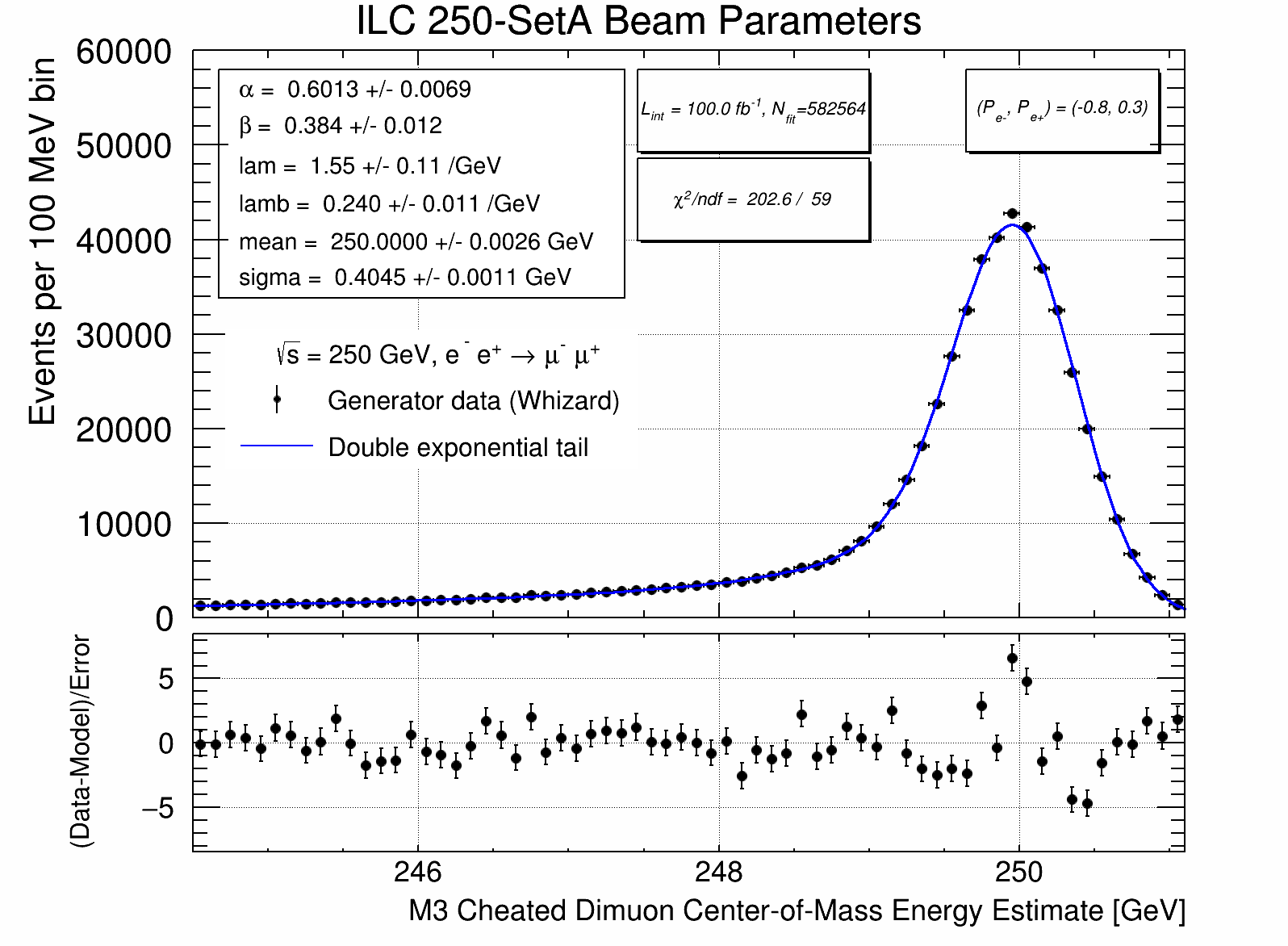}
\caption[]{\small \sl 
Generator level $\sqrtsp$ distribution evaluated with generator-level muons 
for $\eemmg$ events with super-imposed 6-parameter 
fit where the true event-by-event $M_{3}$ value is used.
}
\label{fig:sqrtsp-cheatM3}
\end{figure}

Figure~\ref{fig:sqrtsp} shows the generator level 
distribution of $\sqrtsp$ evaluated with case 2 
together with the same double exponential tail fit 
previously used for the true $\sqrt{s}$ distribution. 
Figure~\ref{fig:sqrtspcheated} shows the distribution 
of $\sqrtsp$ evaluated 
with case 3 where one assumes one knows the true value of $\Ediff$. 
It is noteworthy that the degradation 
of the peak width under the $\Ediff=0$ assumption is recuperated by 
using the correct value for $\Ediff$. Similarly, 
Figure~\ref{fig:sqrtsp-cheatM3} shows the distribution of $\sqrtsp$ evaluated with 
the usual equal beam energies assumption, but in this case 
imposing the true value of $M_{3}$ rather than the assumed value of zero. 
In this case the peak width is very similar to case 2 (3\% smaller) indicating 
that the $M_{3}=0$ assumption by itself does not degrade the peak width substantially. 
However, one can see that significantly more events, that were previously undermeasured, 
are now found in the peak region.  

Table~\ref{tab-example-all} lists 
various kinematic quantities 
at generator-level 
for six illustrative events. 
As expected, when $M_{3}$ is small, 
the $\Ediff$ cheated $\sqrtsp$ estimate 
is identical to the true $\sqrt{s}$ (Events 1,2,4,5,6).
Additionally, when $\Ediff$ is 
also small, the $\sqrtsp$ estimate 
agrees rather well with 
the true $\sqrt{s}$ for 
both radiative-return Z type 
events (Event 1) and full-energy 
events (Event 4). Events 2, 5, 
and 6 have large values of $|\Ediff|$ 
leading to large deviations in $\sqrtsp$ 
from $\sqrt{s}$. In event 2, where 
there is no photon, 
the $\sqrtsp$ estimator 
overestimates by 
about $2 |\Ediff|$ 
over $M_{12}$ by adding a 
fictitious 10~GeV photon in 
the electron direction.
While in events 5 and 6, the $\sqrtsp$ 
estimator either 
largely overestimates 
or largely underestimates $\sqrt{s}$ 
depending on whether 
the inferred photonic system 
is in the same or opposite 
hemisphere to the direction of 
the initial state $\Ediff$
induced boost.
The biases are obviously  
big for these two individual events, 
but the main effect 
is a tendency to broaden the 
estimated $\sqrt{s}$  
since radiation of ISR photons 
should occur symmetrically for 
the electron and positron beams.
Lastly, event 3 has a 
massive photonic system, 
and a substantial $|\Ediff|$ 
which is then problematic for 
both the $M_3=0$ 
and $\Ediff=0$ assumptions.

\begin{table}[hbt]
\begin{center}
\begin{tabular}{|c|r|r|r|r|r|r|}
\hline 
     Event                      &   1 &       2 &       3 &     4 &     5 &     6    \\  \hline
     $\Ebm$                     & 125.34 &   114.55 &   125.32 & 124.87 & 124.75 & 122.77    \\  
     $\Ebp$                     & 124.82 &  124.64 &   121.08 & 124.49 & 116.24 & 110.12    \\ 
     $\Ediff$                   &  $+0.26$ &   $-5.04$ &    $+2.12$ &  $+0.19$ &  $+4.26$ &  $+6.33$    \\     
     $M_{12}$                   &  92.55 &   238.97 &    94.62 & 249.30 &  82.34 &  92.26    \\
     $p_{12}$                   &  108.41 &    10.22 &   104.74 &   1.73 & 101.66 & 105.43    \\
     $p_{12}^{x}$               &  $+18.82$ &    $+1.67$ &    $+1.25$ &  $+1.70$ &$+0.92$ &$+1.03$    \\ 
     $p_{12}^{y}$               &  $-14.54$ &     0.00 &    $+0.21$ &  $-0.01$ & 0.00 &$-0.25$    \\             
     $p_{12}^{z}$               & $+105.77$ &   $-10.08$ &  $+104.73$ &  $+0.35$ &$-101.65$ &$+105.43$  \\          
     $p_{3}$                    &  107.62 &     0.00 &   100.49 &   0.06 & 110.17 &  92.78    \\
     $M_{3}$                    &  0.00 &     0.00 &    31.27 &   0.00 &   0.55 &   0.00    \\   \hline   
    $\sqrt{s}$                  & 250.15  &   238.97 &   246.35 & 249.35 & 240.84 & 232.53    \\  \hline  
     $E_{12}^{*}$ ($\beta_{x}$)              &  142.41 &  239.18  &   141.15 &  249.30 & 130.82  & 140.10   \\ 
     $p_{12}^{*}$ ($\beta_{x}$)              &  108.24 &   10.08  &   104.73 &    0.35 & 101.65  & 105.43   \\      
    $\sqrtsp$                   &   250.65 &  249.26 &   245.88 & 249.65 & 232.47 & 245.53    \\ \hline         
     $E_{12}^{*}$ ($\beta_{x}, \beta_{z}$)     &  142.20 &  238.97  &   139.36 &  249.30 & 134.49  & 134.57   \\ 
     $p_{12}^{*}$  ($\beta_{x}, \beta_{z}$)    &  107.96 &    0.00  &   102.32 &    0.06 & 106.34  &  97.96   \\      
    $\sqrtsp$ (true $\Ediff$)   &   250.15 &  238.97 &   241.60 & 249.35 & 240.84 & 232.53    \\ \hline         
    $\sqrtsp$ (true $M_{3}$)    &   250.65 & 249.26 &   250.45 & 249.65 & 232.47 & 245.53    \\  \hline  

\end{tabular}
\end{center}
\caption{Generator-level kinematic quantities (GeV) for illustrative events. 
Energies and momenta are given in the lab. frame except 
for $E_{12}^{*}$ and $p_{12}^{*} = \sqrt{(E_{12}^{*})^{2} - M_{12}^{2}}$ 
which are the dimuon energy and momentum after boosting according to either the 
Case 2 $\sqrtsp$ assumptions ($\beta_x$) 
or the correct Case 3 boost ($\beta_x$, $\beta_{z}$) 
as discussed in~\ref{sec:alt}.}
\label{tab-example-all}
\end{table}

\section{Dimuon event selection and reconstruction}
\label{sec:analysis}
The event selection used is very simple given that 
it is not expected that backgrounds will be quantitatively important. 
Detector acceptance is the main limiting factor for overall efficiency 
and this is particularly 
true for reconstructing $\eemmg$ events where the dimuon mass is very low.
Quality of momentum reconstruction is very important for reconstructing $\sqrt{s}$ 
with this method and this is a strong function of polar angle. 
The generated $\eemmg$ events are from two samples one 
using $\Pel = -1.0, \Ppos = + 1.0$ (LR) with a cross-section of 11.18~pb and one 
using $\Pel =  +1.0, \Ppos = - 1.0$ (RL) with a cross-section of 8.80~pb. These are mixed 
together to form partially polarized (80\%/30\%) mixtures for the four helicity 
combinations (LR, RL, LL, RR) associated with the 
standard ILC running scenario with integrated luminosity 
fractions of (45\%, 45\%, 5\%, 5\%) respectively.

\begin{figure}[!htbp]
\centering
\includegraphics[height=0.4\textheight]{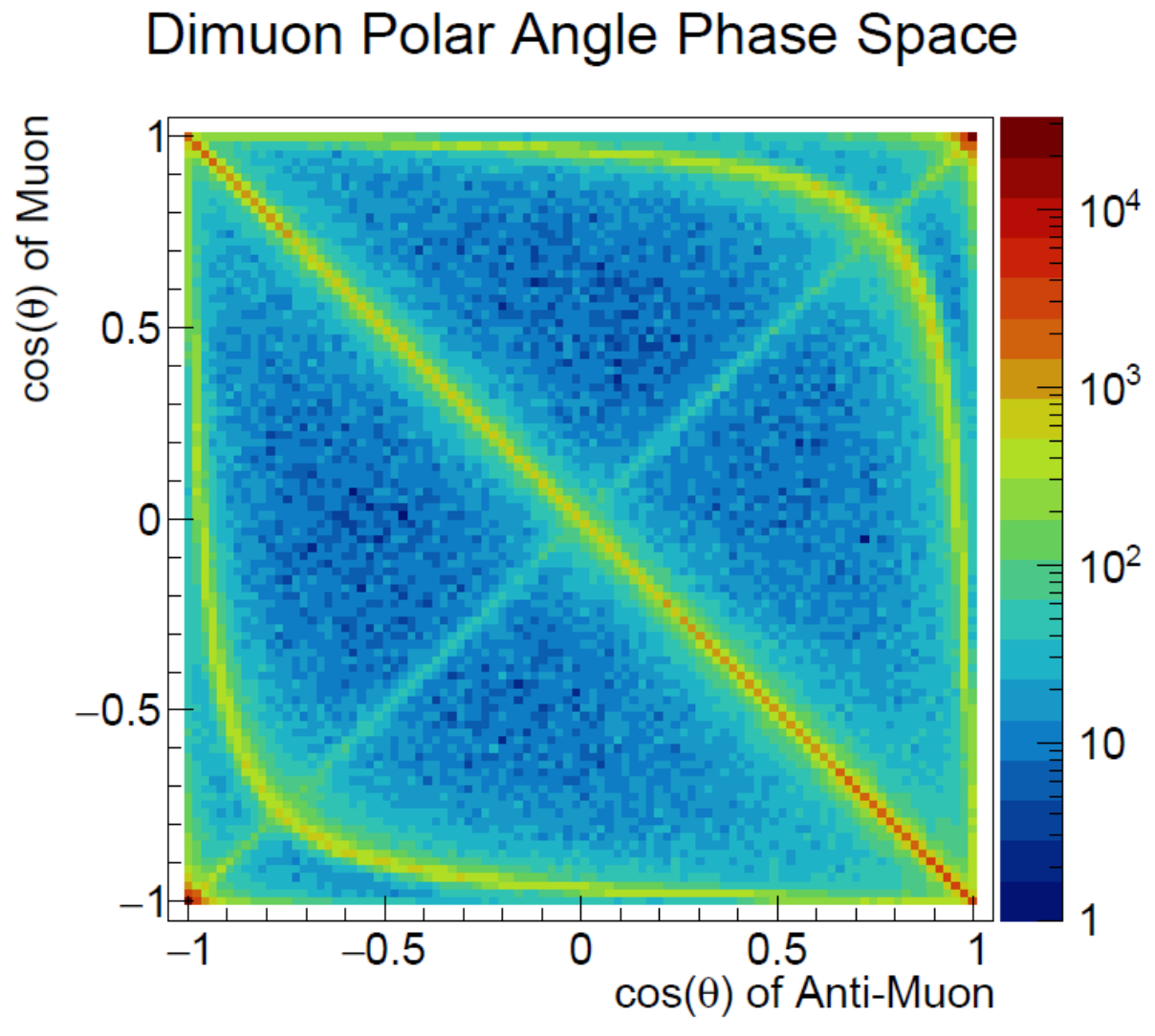}
\caption[]{\small \sl 
Generator level distribution of the cosines of the 
muon polar angles for $\sqrt{s}=250$~GeV for $\eemmg$ 
with $P(\electron)=-0.8$ and $P(\positron)=+0.3$.
}
\label{fig:angles}
\end{figure}

Figure~\ref{fig:angles} illustrates the distribution of the cosine 
of the muon polar angles at generator level for $\sqrt{s}=250$~GeV. The events 
include dimuons with $M_{\mu\mu}$ near ``full-energy'' of 250~GeV, 
the radiative-return to the Z events with $M_{\mu\mu} \approx \mZ$ and radiative events 
with much lower $M_{\mu\mu}$ extending all the way to the threshold at $2 m_{\mu}$.
The full-energy events populate the collinear region of this figure where the two muons tend 
to be back-to-back in space (the descending diagonal). 
The radiative-return to the Z events populate the two relatively narrow 
hyperbola regions (the width is dictated partly by $\GZ$). 
The lower-left/upper-right hyperbola regions 
correspond to emission of an approximately 108~GeV ISR photon 
by the electron/positron beam respectively. Additionally 
there are populated regions at the extreme lower-left and upper-right arising 
from radiative-return to very low mass where the dimuon system tends 
to be at very forward angle balancing an essentially beam energy ISR photon emitted 
preferentially close to the beam axis. Lastly the events on the ascending diagonal 
are (much rarer) events with rather low $M_{\mu\mu}$ where the beam energy photon 
is scattered at large polar angle, resulting in two muons of low mass but 
with dimuon energy close to the beam energy going in very similar directions.

The study uses the latest ILD full simulation and reconstruction samples based on the 
ILC accelerator model at $\sqrt{s}=250$~GeV 
for beam energy spread and beamstrahlung and 
the IDR-L detector model described in~\cite{ILDConceptGroup:2020sfq}.
The event selection criteria applied are:
\begin{itemize}
 \item Require exactly two reconstructed muon particle flow objects and require that they have 
 opposite charge (84.1\% pass).
 \item Require that the calculated uncertainty on $\sqrtsp$ computed from the 
 two muons be less than 0.8\% of the nominal center-of-mass energy of 250~GeV (72.2\% pass). 
 This is based on propagating the track error matrices.
 \item Require that the two muons pass 
 a common vertex fit with p-value exceeding 1\% (70.4\% pass).
 This reduces non-prompt backgrounds such as events where one 
of the muons is from a muonically decaying tau lepton. It also likely makes 
the $\sqrtsp$ estimate more robust, and facilitates future studies 
of primary vertex position dependence of the luminosity spectrum. 
 \end{itemize}
 
 Given above in parentheses are the fractions of LR mixture events surviving these sequential criteria.
 The reconstruction quality in terms of the 
 calculated $\sqrtsp$ uncertainty as a fraction of 
 the nominal center-of-mass energy is used to define three dimuon event quality 
 categories:  gold, silver, and bronze 
 for $<0.15$\%, [0.15\%, 0.30\%], and [0.30\%, 0.80\%] respectively.

The overall efficiency including acceptance effects to pass all criteria 
is about 69.2\% averaged 
over the helicity combinations. The (background) 
efficiency for selecting $\ee \to \tau^{+} \tau^{-} (\gamma)$ events 
is small (about 0.15\%). 
A detailed background study is beyond the scope of the current work.

\begin{figure}[!htbp]
\centering
\includegraphics[height=0.4\textheight]{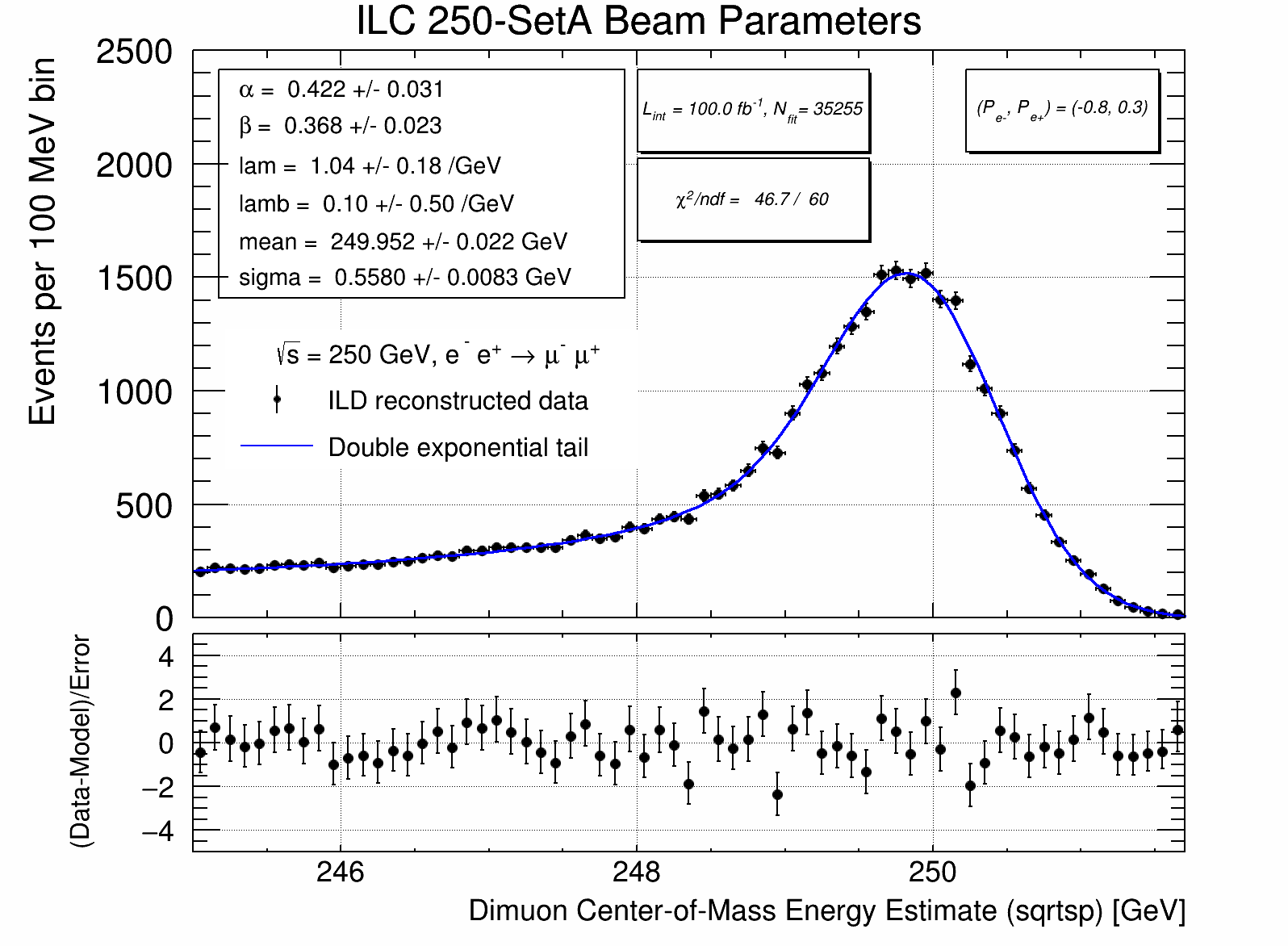}
\caption[]{\small \sl 
Reconstruction level $\sqrtsp$ distribution evaluated with muon 
particle flow objects for $\eemmg$ events 
with super-imposed 6-parameter fit where the dimuon $\sqrtsp$ resolution 
satisfies the gold quality criterion.
}
\label{fig:sqrtspgold}
\end{figure}

\begin{figure}[!htbp]
\centering
\includegraphics[height=0.4\textheight]{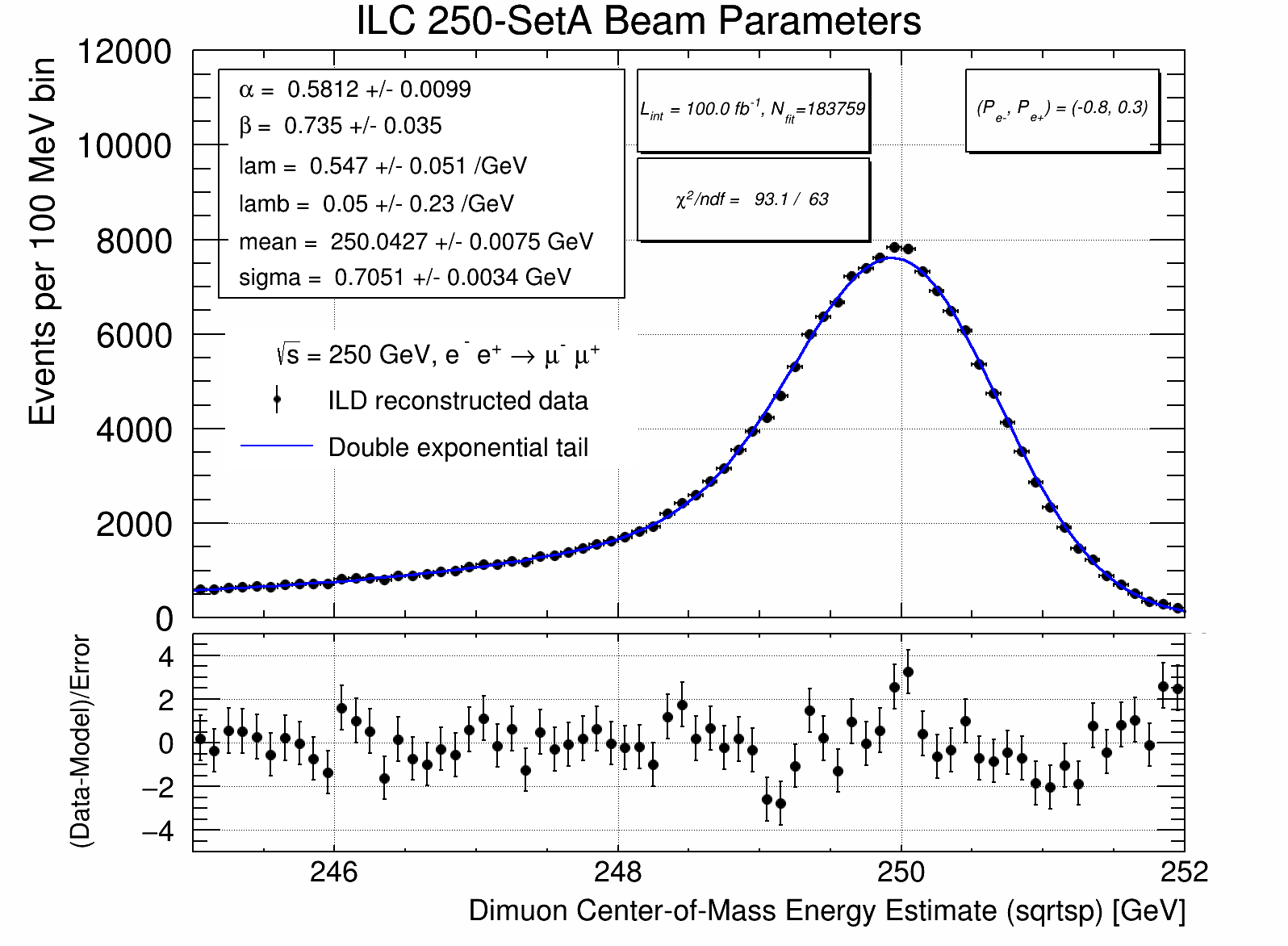}
\caption[]{\small \sl 
Reconstruction level $\sqrtsp$ distribution evaluated with muon 
particle flow objects for $\eemmg$ events 
with super-imposed 6-parameter fit where the dimuon $\sqrtsp$ resolution 
satisfies the silver quality criterion.
}
\label{fig:sqrtspsilver}
\end{figure}

\begin{figure}[!htbp]
\centering
\includegraphics[height=0.4\textheight]{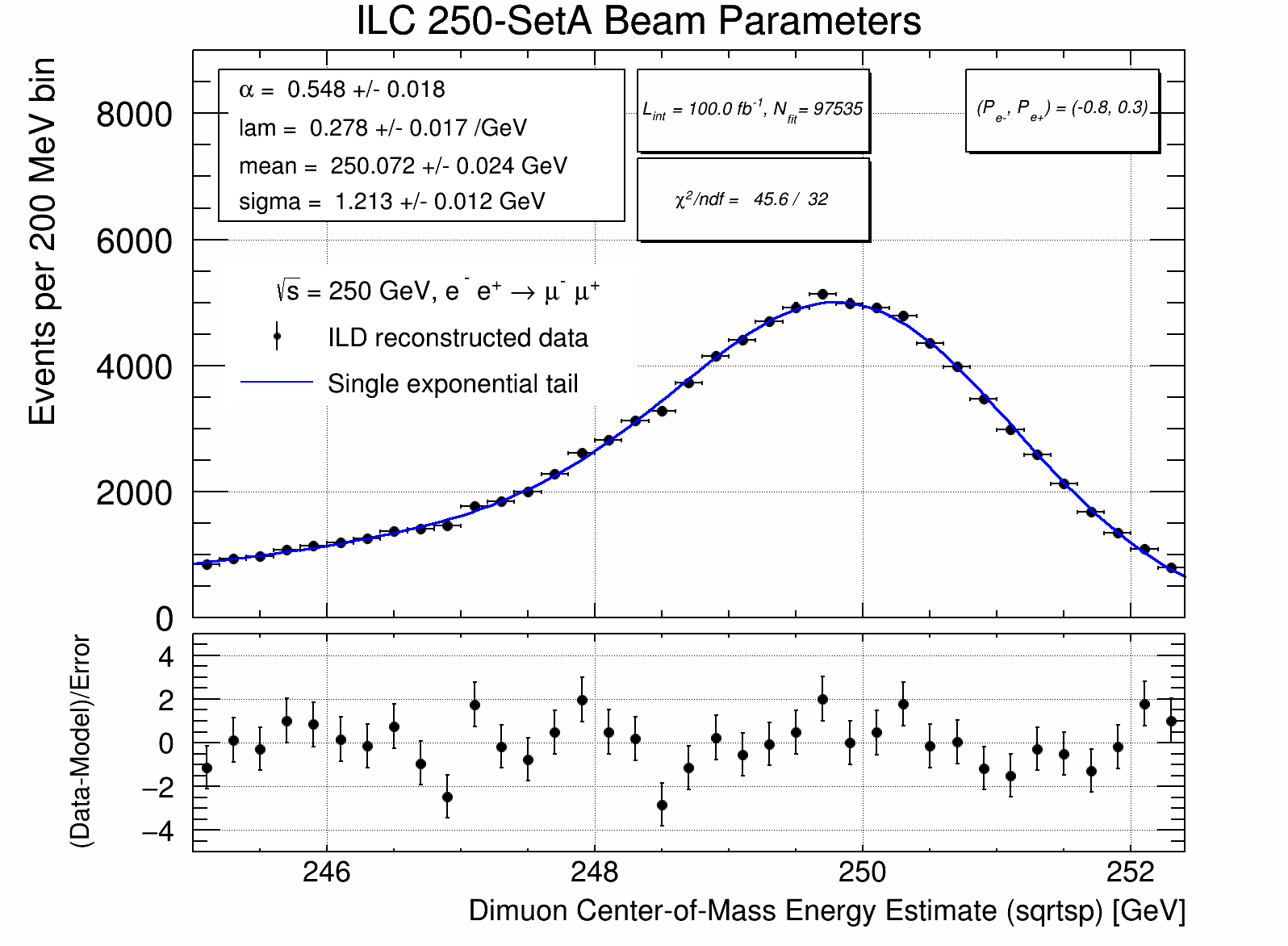}
\caption[]{\small \sl 
Reconstruction level $\sqrtsp$ distribution evaluated with muon 
particle flow objects for $\eemmg$ events 
with super-imposed 6-parameter fit where the dimuon $\sqrtsp$ resolution 
satisfies the bronze quality criterion. In this case a single 
exponential tail component is sufficient.
}
\label{fig:sqrtspbronze}
\end{figure}

The reconstructed $\sqrtsp$ distributions are shown 
in Figures~\ref{fig:sqrtspgold},~\ref{fig:sqrtspsilver}, and \ref{fig:sqrtspbronze} for 
the gold, silver, and bronze categories, respectively. 
The empirical fit model (double exponential tail) 
works reasonably well; it manages to accommodate the combined effects from energy spread, beamstrahlung, detector resolution, ISR, and FSR.
As expected the widths of the peaks degrade from gold to silver to bronze. 
A more refined analysis would try to exploit this per event 
resolution information at the event-to-event level, but for now we 
confine ourselves to estimates of the precision on the 
center-of-mass energy scale from these three separate categories.
These are shown in Table~\ref{tab:results} for the four different data-sets 
associated with the 
ILC standard run plan at $\sqrt{s}=250$~GeV as 
well as for unpolarized beams for comparison. 
Here we have fixed the shape parameters of 
the fits (based on the best fit) and only 
fit for the peak location parameter. 
We have done similar studies 
using the asymmetric Crystal Ball fit model and 
find essentially the same results 
for statistical precision, but appreciably 
different values of $\mu$ as may be expected. 
Both fit models are described in~\ref{app:fits}. Overall the statistical uncertainty associated 
with the 2 $\invab$ dataset is 1.9~ppm on the center-of-mass energy scale.

\begin{table}[!htbp]
\begin{center}
\begin{tabular}{|c|c|c|c|r|r|r|}
\hline
$L_\mathrm{int}$ [ab$^{-1}$] & Poln [\%] & Efficiency [\%]& Gold & Silver & Bronze & All categories  \\ \hline
2.0  &  $  0,  0$ & 69.3   & 5.1  & 2.4 & 6.1 & 2.1 \\ \hline
0.9  &  $-80,+30$ & 70.4 & 6.4 & 3.1 & 7.7 & 2.6\\
0.9  &  $+80,-30$ & 68.0 & 7.5 & 3.4 & 8.7 & 2.9\\
0.1  &  $-80,-30$ & 70.1 & 25   & 12  & 30 & 10 \\
0.1  &  $+80,+30$ & 68.3 & 28   & 13  & 33 & 11\\  \hline
2.0  &  Combined  &  -   & 4.7  & 2.2 & 5.6 & {\bf 1.9} \\ \hline

\end{tabular}
\end{center}
\caption{
Fractional statistical uncertainties in parts per million (ppm) on the mean parameter, $\mu$, of 
the Gaussian peak component of the measured $\sqrtsp$ distribution 
for $\sqrt{s}=250$~GeV ILC running (4 data-sets with different beam polarizations) 
from fits to the gold, silver, and bronze categories of dimuons. These are 1-parameter fits where 
the model shape 
parameters are fixed to their best fit values.
For gold and silver categories, the fit uses the 
double exponential tail function, while for bronze it 
uses the single exponential tail function. Fit ranges and binning are as 
indicated in Figures~\ref{fig:sqrtspgold}, \ref{fig:sqrtspsilver}, and \ref{fig:sqrtspbronze}. 
The all categories column shows the
weighted statistical uncertainty assuming that 
all data-set columns are used in each weighted average.
The last row shows the combination of the four 80\%/30\% polarization 
data-sets. Also shown for comparison in the first row is the result for unpolarized beams.
}
\label{tab:results}
\end{table}

We justify focusing for now on this seemingly artificial single parameter fit 
because we expect the shape parameters to be much better measured using Bhabhas in the 
luminosity spectrum measurement, and they should also be constrained to some extent by 
accelerator diagnostics measurements. 
The main goal of the dimuon studies is to pin down the absolute center-of-mass energy scale that 
is of prime importance from the physics perspective. It is also expected that  
the shape sensitivity can be improved, (and the correlations on $\mu$ reduced), 
by widening the fit range.

Fits have also been performed where the peak mean and width are fitted with 
the other shape parameters fixed, leading to at most a 9\% degradation 
in the precision of the scale parameter, $\mu$. 
It is expected that by applying $\sqrtsp$ to Bhabha events and 
using in this case the tracker-based electron momentum measurement 
that a substantial increase in precision should also be feasible\footnote{ 
Note that the tracker electron momentum scale will likely be more difficult to pin down than that of muons.}.

\begin{table}[!htbp]
\begin{center}
\begin{tabular}{|c|c|c|r|r|}
\hline
$L_\mathrm{int}$ [ab$^{-1}$] & Poln [\%] & Gold & Silver & Bronze  \\ \hline
0.1  &  $-80,+30$ & $249.9273 \pm 0.0048$ & $250.0422 \pm 0.0023$ &  $250.0715 \pm  0.0058$ \\
0.1  &  $+80,-30$ & $249.9335 \pm 0.0056$ & $250.0541 \pm 0.0026$ &  $250.0972 \pm  0.0066$ \\  \hline
\end{tabular}
\end{center}
\caption{
Results of the 1-parameter fits to the reconstructed $\sqrtsp$ distribution 
for the $\mu$ parameter in units of GeV for the three different categories 
for the two main helicity mixtures. This uses the same integrated luminosity 
as the figures (100~$\invfb$) rather than scaling these numbers to the ILC standard running scenario 
as done in Table~\ref{tab:results}.
}
\label{tab:results2}
\end{table}

We also show in Table~\ref{tab:results2} the fitted $\mu$ values for six of the fits. 
It is noteworthy that the fitted $\mu$ values depend markedly on dimuon reconstruction 
quality. We attribute 
some of this to the expectation that high-mass events with little missing momentum are liable to 
be over-measured as a result of momentum resolution. It is also the case 
that the relative fractions 
of radiative-return to the Z and high-energy events is a strong function 
of the $\sqrtsp$ measurement uncertainty. It is expected that constrained kinematic 
fitting may assist with some of these issues.

\begin{table}[!htbp]
\begin{center}
\begin{tabular}{|c|c|r|r|}
\hline
$M_{\mumu}$ range [GeV] & $\mu(\sqrt{s})$ [GeV]   & $\mu(\sqrtsp)$ [GeV] & $\mu(\sqrtsp)$ -  $\mu(\sqrt{s})$ [MeV]  \\ \hline
$M > 150$         & $249.9792 \pm 0.0011$   & $250.0337 \pm 0.0013$   &  $ +54.5 \pm 1.7 $ \\
$50 < M < 150$    & $249.9813 \pm 0.0010$   & $249.9602 \pm 0.0017$   &  $ -21.1 \pm 2.0$ \\ 
$M < 50$          & $249.9871 \pm 0.0015$   & $249.9633 \pm 0.0028$   &  $ -23.8 \pm 3.2 $ \\ \hline
All               & $249.9816 \pm 0.0008$ & $250.0014 \pm 0.0010$ &  $ + 19.8 \pm  1.2 $ \\ 
\hline
\end{tabular}
\end{center}
\caption{
Results of the 1-parameter fits for the $\mu$ parameter to the generator-level 
distributions of $\sqrt{s}$ and $\sqrtsp$ for three different dimuon mass ranges  
for the 80\%/30\% LR helicity mixture. The statistical uncertainties of these tests 
reflect an integrated luminosity of 100~$\invfb$. 
The last column gives the difference in MeV of the fit parameters for 
the two distributions.
}
\label{tab:results3}
\end{table}

To address this we also tabulate in Table~\ref{tab:results3} the results of the 
fits to the $\sqrt{s}$ and $\sqrtsp$ distributions at generator level 
for three dimuon mass ranges. One sees that even without detector 
resolution there is a tendency for high mass events to be over-measured, and for lower mass events 
to be under-measured. 
Looking back and comparing Figure~\ref{fig:sqrtsp} with Figure~\ref{fig:sqrtspcheated}, 
it appears that indeed the $\Ediff = 0$ assumption in the $\sqrtsp$ estimate on average 
induces a $+40$~MeV upward shift in the estimate. Furthermore, the peak region 
of Figure~\ref{fig:sqrtsp} shows some evidence of substructure that may be 
related to these observations.

\section{Outlook and future work}
\label{sec:outlook}
There are a number of promising topics that are very relevant to fully 
realizing the potential of the methods discussed here. Some are already investigated 
to an extent and should be written up in due course and others are appropriate 
for future work.
Some in the latter category need collaborative developments with beam simulations, 
physics generators, and detector simulations. 
These topics include:
\begin{enumerate}
    \item Constrained kinematic fits. For example one can test 
          the consistency with the pure 2-body hypothesis of $\ee \to \mumu$ while 
          fitting for the two unmeasured parameters of $\Eave$ and $\Ediff$, 
          and also perform fits with the $\ee \to \mumu \gamma \;$ hypothesis.
    \item Extending the techniques to the $\ee \to \ee$ channel.
    \item Exploiting fully events with detected photons.
    \item Implementing complete end-to-end measurement scheme and 
          understand how best to use different kinematic regimes 
          and correct/mitigate observed biases.
    \item Characterizing better the intrinsic limitations associated with 
          beam energy spread, beamstrahlung, ISR, FSR, backgrounds, and detector acceptance and resolution. This includes studies with more specialized physics event generators 
          such as KKMCee~\cite{Jadach:2022mbe}.         
    \item Tracker momentum scale studies using $\Jmumu$, $\kspipi$, $\Lppi$. One of us 
         has  developed 
         further the technique advocated in~\cite{Rod2020} based on 
          the Armenteros-Podolanski~\cite{AP-paper} construction. 
          A novel aspect of this approach is that one can aspire to simultaneously 
          improve the measurements of the $\KShort$ and $\Lambda$ masses together with the momentum scale 
          given that the masses of their decay products are very well known. 
          A preliminary conceptual study reported in~\cite{LCWS21} found a statistical uncertainty of 2.5~ppm on the momentum scale per 10~M hadronically decaying Z events.
    \item Understanding the relative merit of dimuons for luminosity spectrum determination 
          compared with Bhabhas and integrating both techniques in a global analysis.
    \item Characterizing further the scope for measuring 
          accelerator parameters such as the crossing angle 
          and beamstrahlung-induced correlations including the observed dependence of 
          the beam energy spectrum on the longitudinal collision vertex. The latter has been shown to be 
          easily measurable with vertex fits in $\eemmg$ events.
\end{enumerate}

\section{Conclusions}
\label{sec:concl}
We have discussed and developed a dimuon based estimator, denoted $\sqrtsp$,  
for determining the center-of-mass energy at a future $\ee$ Higgs-factory collider 
like ILC. It relies on the precision 
measurement of muon momenta, and we have placed this method in the context of established methods.
The underlying statistical precision at ILC at $\sqrt{s}=250$~GeV 
is 1.9 ppm for a 2.0 $\invab$ polarized dataset 
based on full simulation studies with the ILD detector concept. 
This statistical precision is an order of magnitude better than the ``angles method'' using muons.
The $\sqrtsp$ method gives a resolution almost commensurate with the beam energy spread, and 
can work for all masses of the dimuon not just $\mathrm{Z} \gamma$ events or full energy events, 
and can be applied at any $\sqrt{s}$ including at the Z pole and WW threshold.
Realizing this great potential can lead to marked advances in the measurement of the 
masses of the W and Z boson, but will need more study of the evident 
underlying systematic issues that likely can be resolved.
The ultimate utility will depend largely on how well one can calibrate and maintain 
the tracker momentum scale.

\section*{Acknowledgments}
We acknowledge earlier contributions of Jadranka Sekaric to the fit modeling. 
We thank Tim Barklow and Alberto Ruiz 
for helpful comments.
We would like to thank the LCC generator working group and 
the ILD software working group for providing the 
simulation and reconstruction tools and producing the Monte Carlo samples 
used in this study.

The work of Graham Wilson is partially supported by 
the US National Science Foundation under award NSF 2013007. 
This work benefited from use of the HPC facilities operated by 
the Center for Research Computing at the University of Kansas. 
This work has also benefited from computing services provided by 
the ILC Virtual Organization, supported by the national resource 
providers of the EGI Federation and the Open Science GRID.


\appendix
\section{Alternative formulation}
\label{sec:alt}
Under the assumption of equal beam energies ($\Ediff=0$), one 
can also write the underlying equations directly 
in terms of a quadratic in $\sqrt{s}$. 
This started under the same assumptions
of $p_y$ and $p_{z}$ balance,
and again with no requirements on the directions of the 
rest-of-the-event particles\footnote{The 
related approach of equations 8.1 to 8.4 in~\cite{Blondel:2019jmp} posits that 
one of the beams radiates a collinear ISR photon.}.
This is as case 2 in Section~\ref{sec:sqrtsp} but, 
more explicitly, the assumption of equal beam energies leads to 
\begin{equation}  \sqrt{s} = 2 \; \Eave \; \ca \; \text{,} \end{equation}
and we can then rewrite the basic equations again retaining a potentially massive rest-of-the-event hypothesis as
\begin{equation} E_{1} + E_{2} + \sqrt{p_{3}^2 + M_{3}^{2}} = \frac{\sqrt{s}}{\ca} \; \text{,}   \end{equation}
\begin{equation} \ponebf + \ptwobf + \pthreebf = ( \sqrt{s} \; \ta, 0 , 0 ) 
= \pinibf \; \text{.}\end{equation} 

Eliminating $\pthreebf$ from these two equations leads to a 
quadratic in $\sqrt{s}$, with known or measured coefficients, and assumed $M_{3}$, 
namely,

\begin{equation} s \; + \; \frac{2[ p_{12}^{x} \sa - E_{12}]} {\ca} \sqrt{s} \; + \; (M_{12}^{2} - M_{3}^{2}) = 0 \; \text{.}\end{equation}

We started with four equations with five unknowns which can be solved 
for $\sqrt{s}$ if $M_{3}$ is specified and the dimuon system is measured.
Examining the discriminant of this quadratic equation, we have been able to 
show that it must be non-negative for the particular case of $M_{3}=0$. 

Looking more closely, the positive solution of the quadratic
\begin{equation}
\sqrt{s}_{+} = \frac{E_{12} - p_{12}^{x}  \sa} {\ca} + \sqrt{ \left( \frac{E_{12} - p_{12}^{x}  \sa} {\ca}\right)^{2} + (M_{3}^{2} - M_{12}^{2})  } \; \text{,}
\end{equation}
can be understood physically as consisting of
\begin{equation}
\sqrt{s}_{+} = E_{12}^{*}+ \sqrt{ ( {E_{12}^{*}}^{2} - M_{12}^{2})  + M_{3}^{2}} =  E_{12}^{*}+  \sqrt{ {p_{12}^{*}}^{2}  + M_{3}^{2}} =       E_{12}^{*} + E_{3}^{*} \; \text{,}
\label{eqn:sqrtsplus}
\end{equation}
where in the last step we use $\mathrm{\mathbf{p}}_{12}^{*} = - \mathrm{\mathbf{p}}_{3}^{*}$ and the dimuon energy in the center-of-momentum 
frame, $E_{12}^{*}$, is identified as, 
\begin{equation}
E_{12}^{*} = \frac{E_{12} - p_{12}^{x}  \sa} {\ca} = \gamma ( E_{12} - \beta p_{12}^{x} ) \; \text{,}
\end{equation}
with Lorentz boost factors of $\beta = \sa$, $\gamma=1/\ca$. 
This corresponds to a boost from the lab with crossing angle back 
to the center-of-momentum frame in the negative $x$ direction. 
As expected, the center-of-mass energy is the sum of the energies 
of the constituent components. Furthermore the dimuon and rest-of-the 
event term are as identified. Under the $M_{3}=0$ 
assumption, Equation~\ref{eqn:sqrtsplus} 
is essentially identical to the original naive formulation now 
that quantities are actually evaluated in 
the center-of-momentum frame, namely, 
\begin{equation}
\sqrt{s}_{+} = E_{12}^{*}+   |\mathrm{\mathbf{p}}_{12}^{*}|  \; \text{.}
\label{eqn:c2-epiphany}
\end{equation}

Similarly, one can also show 
for case 3, that the solution 
with the correct choice of $\Ediff$ (for the component of the boost 
along the $z$-axis), leads 
again to Equation~\ref{eqn:c2-epiphany} but in this case with, 
\begin{equation}
E_{12}^{*} = \gamma ( E_{12} - \beta_{x} p_{12}^{x} - \beta_{z} p_{12}^{z}) \; \text{,}
\end{equation}
where $\beta_{x} = \sa$, $\beta_{z} = (\Ediff/\Eave) \ca$ and $\gamma = (1 - \beta_{x}^{2} - \beta_{z}^{2})^{-\frac{1}{2}}$.

\section{Fit models}
\label{app:fits}
We have used two main fit models to characterize observations and 
expected sensitivites to the center-of-mass energy scale parameter.
In earlier work we had focused on the use of 
the well known Crystal Ball function, and we are now preferring to use 
a more physically motivated convolution based parametrization. 
We describe both here.

\subsection{Asymmetric Crystal Ball}
Our Crystal Ball implementation uses the RooCrystalBall probability density function  
from the RooFit framework~\cite{Verkerke:2003ir}. This 
is an asymmetric double-sided Crystal Ball function 
with up to seven parameters. It 
is defined in four piece-wise regions 
of the scaled deviation from the mean parameter 
for the left and right Gaussian. Parameters 1-3 define 
the peak position and shape, and parameters 4-7 are related to the locations in number of standard deviations 
of the power-law transitions, and the exponents of the power-laws.
\begin{itemize}
\item 1. Location parameter ($\mu_{0}$) 
\item 2 and 3. Left/right Gaussian width, $\sigma_{L}$, $\sigma_{R}$
\item 4 and 5. Left/right transition point, $\alpha_{L}$, $\alpha_{R}$ (in units of $\sigma_{L}$, $\sigma_{R}$)
\item 6 and 7. Left/right exponents, $n_{L}$, $n_{R}$ 
\end{itemize}

We used this 7-parameter function for the fit 
to the energy difference distribution of Figure~\ref{fig:ediff} 
where we wanted the flexibility to 
check for asymmetries. In our 5-parameter versions of this function that 
we have also used for most of the 
lossy energy distributions, we have effectively turned off the 
right tail parameters by fixing $n_{R}$ 
and fixing $\alpha_{R}$ to +100$\sigma_{R}$.

\subsection{Double exponential tail}
We essentially used the function described 
by equation 9 in~\cite{Cheng:2016ykr} that was originally 
conceived to describe the convolution of 
a calorimetric energy loss function including retention of the full 
absorption peak with a Gaussian detector resolution 
response function. We chose this function for convenience 
because the convolution is analytically integrable (see equation 8 of Cheng).
This contrasts with the traditional CIRCE function for the tail based on beta distributions described 
in~\cite{Ohl:1996fi} which is thought not to be analytically integrable~\cite{Poss:2013oea} 
and necessitates numerical convolutions.
We implemented this analytic equation for 
the probability density function adjusting for a different convention 
for the fractions in each exponential component.
For completeness, the double exponential tail model we implemented is meant to correspond to the
convolution integral 
\begin{multline}
    f(E; \mu, \sigma, \alpha, \beta, \lambda_{1}, \lambda_{2}) = \int_{0}^{\mu}
     ( \alpha \delta(\Ep - \mu) + (1-\alpha) [ \beta f_{\mathrm{exp}}(\Ep; \lambda_{1}) 
    + \\ (1 -\beta ) f_{\mathrm{exp}}(\Ep; \lambda_{2}) ] ) \; G(E-\Ep; \sigma)  d\Ep \; \text{,}  
    \label{eqn:conv}
\end{multline}
where $E$ is the measured energy variable after convolution with the Gaussian response function, $\Ep$ is the true unconvolved energy variable, 
and the limit in the convolution integral for the true energy goes 
from zero to the full energy,~$\mu$.
The Gaussian peak fraction is $\alpha$ and it is represented with a Dirac delta function that is non-zero 
for true energy, $\mu$. 
The Gaussian response function used for the beam energy spread is
\begin{equation}
    G(E - \Ep ; \sigma) = \frac{1}{\sqrt{2 \pi \sigma^2}} \exp\left( - \frac{(E - \Ep)^{2}} {2 \sigma^2} \right) \; \text{,}
\end{equation}
which smears deviations from the true energy, $\Ep$, of size, $E - \Ep$, with resolution, $\sigma$.
The tail has two exponential components, which themselves are also convolved with the response function.
The first has fraction $(1 - \alpha) \beta$ and constant $\lambda_{1}$, 
and the second has fraction $(1-\alpha) (1-\beta)$ 
with constant $\lambda_{2}$ (these fractions differ from Cheng equation 9).
Rather than implementing the exponential as the distribution of 
energy loss, we followed the prescription in Cheng, where 
the ``exponential'' distribution\footnote{In this case it is an exponentially rising distribution with finite support.} 
appears to be defined as
\begin{equation}
    f_{\mathrm{exp}} (\Ep; \lambda) = \lambda \exp( \lambda \Ep ) \; \text{,}
\end{equation}
where $\lambda$ is a positive parameter, $\Ep$ is defined for $\Ep \in [0, \mu]$, and the maximum 
of the distribution is attained at $\Ep = \mu$. 
In the figures, the parameters labeled as mean, sigma, lam, and lamb are $\mu, \sigma, \lambda_{1}$, and $\lambda_{2}$, respectively. 
This fit function manages to describe the simulated data 
much better than the asymmetric Crystal Ball and we have adopted it as 
our current preferred empirical fit model. 

\subsection{Single exponential tail}
In one of the fits we have removed the second exponential tail 
function corresponding to setting $\beta=1$ 
in Equation~\ref{eqn:conv}.

\end{document}